\documentclass[%
 reprint, twocolumn,
 amsmath,amssymb,
prx,
floatfix,
]{revtex4-1}

\usepackage{graphicx}
\usepackage{xcolor}

\begin{document}
\title{ Laser structured micro-targets generate MeV electron temperature
 at $4 \times 10^{16}$ W/cm$^2$}




\author{Angana Mondal$^{1,2,\dagger}$,  Ratul Sabui$^{2}$, Sheroy Tata$^{1}$, R.M.G.M Trines$^3$, S.V. Rahul$^{2}$, Feiyu Li$^{4}$, Soubhik Sarkar$^{1}$, William Trickey$^{5}$, Rakesh Y. Kumar$^{2}$, Debobrata Rajak$^{2}$, John Pasley$^{5}$, Zhengming Sheng$^{4,6}$, J. Jha$^{1}$, M. Anand$^{2}$, Ram Gopal$^{2}$, A.P.L. Robinson$^3$ and M.Krishnamurthy$^{1,2}$}


\email{mkrism@tifr.res.in}


\affiliation{$^1$Tata Institute of Fundamental Research,Homi Bhabha Road, Navy Nagar, Colaba, Mumbai 400005, INDIA}
\affiliation{$^2$Tata Institute of Fundamental Research, Hyderabad, Gopanpally, Serlingampalli, Telangana, INDIA, 500107}
\affiliation{$^{\dagger}$Currently at ETH Zurich, 8093, Switzerland.}
\affiliation{$^3$Central Laser Facility, Rutherford Appleton Laboratory, Chilton, Didcot OX10 0QX, UK}
\affiliation{ $^4$Department of Physics, University of Strathclyde, Glasgow G4 0NG, UK}
\affiliation{$^5$York Plasma Institute, Department of Physics, University of York, York, YO10 5DD, United Kingdom}
\affiliation{  $^6$Laboratory for Laser Plasmas and School of Physics and Astronomy, Shanghai Jiao Tong University, Shanghai 200240, China}

\date{\today}
\begin{abstract}
\noindent Relativistic temperature electrons higher than 0.5 MeV are generated typically with laser intensities of about 10$^{18}$  W/cm$^{2}$. Their generation with  high repetition rate  lasers that operate at non-relativistic intensities ($\simeq$10$^{16}$  W/cm$^{2}$) is cardinal for the realisation of compact, ultra-short, bench-top electron sources.   New strategies, capable of exploiting  different aspects of laser-plasma interaction, are necessary for reducing the required intensity. We report here, a novel technique of dynamic target structuring of microdroplets, capable of generating 200 keV and 1 MeV electron temperatures at 1/100th of the intensity required by ponderomotive scaling($10^{18}$ W/cm$^2$) to generate relativistic electron temperature. Combining the concepts of  pre-plasma tailoring, optimised scale length  and micro-optics, this method achieves two-plasmon decay boosted electron acceleration with "non-ideal" ultrashort (25 fs) pulses at $4\times10^{16}$ W/cm$^2$, only. With shot repeatability at kHz, this precise in-situ targetry produces directed, imaging quality beam-like electron emission up to 6 MeV with milli-joule class lasers, that can be transformational for time-resolved, microscopic studies in all fields of science.
\end{abstract}
\maketitle

\section{Introduction}
\noindent Intense ultrashort laser pulse interactions transfer energy to electrons in matter by the generation of plasma waves. High electric field gradients generated in a localised plasma makes these efficient low emittance high brightness source of electrons and secondary radiation\cite{hbright1, hbright2, hbright3, hbright4}.
Enhancement of these field gradients, therefore forms, the cornerstone of achieving such electron distributions with relativistic temperatures. A straight forward approach which has acquired prominence, is to use relativistic intensities for inducing schemes like laser wakefield acceleration~\cite{lwfa} or direct laser acceleration~\cite{DLA}.  In comparison, the ability to generate a hotter dense plasma with reduced laser intensity still remains ellusive and is important not only from a fundamental perspective but also for envisaging application to $\gamma$-radio radiography, cancer therapy, electron microscopy \cite{gamma1, gamma2, gamma3, cancer1, cancer2, electron1, electron2} that impact a wider gamut of science.

\noindent Enhanced transfer of energy from the light fields to the plasma waves  entails devising optimal parameters for the laser and the  matter  that absorbs the laser pulses. While noteworthy target modification experiments have been performed over many years  \cite{struct1, struct2, struct3, struct4, struct5},  none have efficiently achieved relativistic electron temperatures even with moderate intensity small scale lasers \cite{tg1, anand1}. In this context, we present a new strategy that combines  different features of plasma physics into one morphologically structured, size-limited droplet target, exposed to two collinear mJ laser pulses that produces a directed electron emission with T$_{hot}$ (hot electron temperature) components of 200 keV and 1 MeV even at only 4 $\times$ 10$^{16}$ W/cm$^2$. 

\noindent Light penetrates in a plasma till critical density layer (n$_c$) and deposits energy dominantly where the laser plasma frequency ($\omega_p$) matches  the light frequency $\omega$. When the drive laser intensity is high, it can non-linearly excite plasma waves of $\omega_{p}/2$ at a quarter critical density (n$_{c}/4$) in two-plasmon decay (TPD) mechanism\cite{TPD}.  The plasmons generated, $\omega_{p}/2$,  continue to interact with the laser which propagates upto n$_c$. The signature of the TPD is often seen via  Stimulated Raman scattered signal  (SRS) that is formed at $3\omega_{p}/2$ \cite{TPD, Xray}. Conventionally, such processes require long pulse(ps to ns) lasers with high energy(J to kJ)(\cite{TPD, TPD2}) that can simultaneously generate a spatially extended pre-plasma and attain the threshold intensities for parametric instability onset~\cite{thresh}.
Femtosecond  (fs) pulses with steeper electron density profile are not optimal to take advantage of these effects  and  hot electron temperature is only about 150 keV  even for relativistic intensities of $\sim 10^{18} W/cm^{2}$ \cite{prashant}. However, if these effects can be invoked with millijoule (mJ), few fs($\leq$ 100 fs) laser pulses, it would be possible to achieve unprecedented hot electron temperatures   at sub-relativistic intensities and kHz repetition rates.  Single shot electron/X-ray radiography that are domain of the petawatt class lasers would  be amenable even with modest laser systems. This papers brings out experiments corroborated with simulations towards such a strategy. 

\noindent This methodology is on based  three aspects: i) dynamically shaped critical density surface of optimal length and concavity, ii) appropriate  plasma density gradient about the concave critical density surface iii) a mesoscopic liquid droplet target suspended in vacuum that is used to generate features (i) and (ii).  A collinear pre-pulse sets up hydrodynamic heating and expansion of the convex liquid surface to appropriate concave  surface. A  dynamically cavitated concave liquid drop structure is shown to be formed  both from experimental shadowgrams and hydrodynamic simulations~\cite{h2d}. The main pulse incident at the apex of  the concave surface generate a concave critical density surface. Evaporating liquid provides the low density background which is ionised  by the focusing laser pulse even at  $10^{14} W/cm^{2}$ and provides a plasma gradient for generation of the plasma waves.  Due to longer effective density scale lengths, oblique incidence TPD is driven harder in a plasma than normal incidence TPD~\cite{oblique1, oblique2, oblique3}, reducing the instability onset threshold.
The combined effect of the cup structure and the pre-plasma gradient successfully drives TPD and electron acceleration with T$_{hot}$ components of 200 keV and 1 MeV even at a low intensity of  4 $\times$ 10$^{16}$ W/cm$^2$.  

\noindent Experiments well corroborated by detailed through 2D particle-in-cell(PIC) simulations (SMILEI and EPOCH)~\cite{PIC, EPOCH} show that the absence of  the low density gradient is detrimental to hot electron generation. The two-plasmon generation is shown to be directly controlled by the laser polarisation and yields  a directed twin beams of hot electrons  symmetrical to the laser incidence.  A direct observation of the 3$\omega$/2 emission establishes the TPD mechanism responsible for  electron energies as large as 6MeV even with the "non-ideal", low energy (2 mJ), ultra-short(25 fs) pulses. A detailed assessment is also made of how the electron beam properties change with the different the laser-plasma parameters.

\noindent The present report not only unravels the acceleration mechanism, but also demonstrates the applicability of this directed, high resolution, electron/ X-ray source in radiography and tomography applications. The 200 keV component is significantly dominant in the electron spectrum and is demonstrated to be usable for  single-shot imaging even though it is operated at a kHz repetition rate. Simulations are used to show that  the 200 keV and 1 MeV energy electrons components would have pulse durations of 50 fs and 5 fs, respectively. As a completion to the discussion, we motivate with preliminary results that indicate that the methodolgy is scalable to MHz repetition rate with the currently available state-of-the-art lasers\cite{MHz}.  Thus  moderate intensity, mJ lasers can generate relativistic electron temperature plasma and would be  indispensable stroboscope for ultrafast research in science, engineering and medicine.\\

\section{Results}
\subsection{Experimental Approach}
\noindent A schematic, of the experimental set-up is shown in Fig. 1(A). A detailed description of the detection methods are presented in the Appendix. In brief, the experiments are performed using a 2 mJ, 800 nm, 1 kHz Ti-Sapphire laser system generating 25 fs pulses. The corresponding on-target intensity being $4 \times 10^{16}$ W/cm$^2 $, when focussed to a beam waist of 11 $\mu m$. The target droplets are produced by fragmenting a 10 $\mu$m, pressurized (20 bar) Methanol jet, to a 1 MHz stream of equally separated 15 $\mu$m droplets with a piezoelectric vibrator. The experiments are performed at an ambient pressure of $ 5 \times 10^{-1}$ mbar. Prior to the main laser pulse interaction, dynamical structures, vital for the relativistic electrons, are generated by focussing a collinear pre-pulse on to the target droplet. The pre-pulse, containing only  $ 5 \% $ of the main pulse energy and arriving 4 ns ahead, is identical in all other respects to the main pulse. The incident laser polarization is fixed with a half-wave plate. The overlap of the droplet and the laser pulses, is monitored by visualization and measurement of the transmitted beam and maximization of the electron energy. Emerging electrons are detected using a magnetic field (0.11 T) bending electron spectrometer (ESM) with LANEX or IP as detectors. \\
\noindent Standard laser-plasma mechanisms predict that a spherical dielectric target of size about 20 times the wavelength and a focal spot size smaller than the target, should be akin to a macroscopic target, possibly with small local field enhancements caused by near field scattering at the spherical interface. On interaction with a 25 fs pulse, such a system is expected to generate an electron temperature of about 19 keV even  at $\sim 8.5 \times$ 10$^{16}$ W/cm$^2$ intensity. Fig. 1(B) shows a comparison of the electron spectrum obtained from 2D-PIC for a spherical target with the experimentally obtained electron spectrum. The simulation parameters are elaborated in section II B (2). For the simple spherical target interaction, the high energy electron yield is seen to fall-off rapidly beyond 300 keV, with a sharp cut-off at about 400 keV. In contrast, our experimental measurements, obtained by electron spectrometry, show a dominant $\sim$ 220 keV temperature  measurable on a shot-to-shot basis (though $\sim$ 8 s of data acquisition is presented for better statistics) using LANEX detectors(Fig. 1(B)) and a $\sim$ 1.3 MeV electron temperature component extending to about 6 MeV(Fig. 1(C)). Due to their weaker yield(few percent of 200 keV component), the 1 MeV component is captured in only long-time acquisition with the IP detectors that have 10$^5$ orders signal detection range as opposed to that of 10$^2$ in LANEX. The observed electron temperatures are also verified with bremsstrahlung X-ray measurements presented in Fig. S1(A). A study of the dominant 200 keV temperature component as a function of laser intensity shows a systematic, non-linear increase in the temperature and yield (Fig. S1(B) and (C)) with the laser intensity (I), following a scaling of I$^{4.5\pm0.3}$.\\ 
\noindent To decipher the underlying mechanism, the angular distribution and the polarization dependence of the emitted electrons are measured. Fig. 2(A), shows the electron emission to be confined in the polarization plane and oriented along two backward directed cones at angles of about $\pm 50^\circ$ with respect to the incident laser. The directed emission of electrons at the aforementioned angles should not be mistaken as resonance absorption(RA) in spherical targets~\cite{reso, resotheo}, since the obtained T$_{hot}$ and its intensity is about 10 and 13 times larger than would be expected from ponderomotive acceleration and RA, respectively. In addition, unlike in RA, the electron angular distribution is also highly tunable with the change in laser intensity, as shown in Fig. 2(D).\\

\onecolumngrid

\begin{figure}[htbp]
\centering
\includegraphics[ width=\columnwidth]{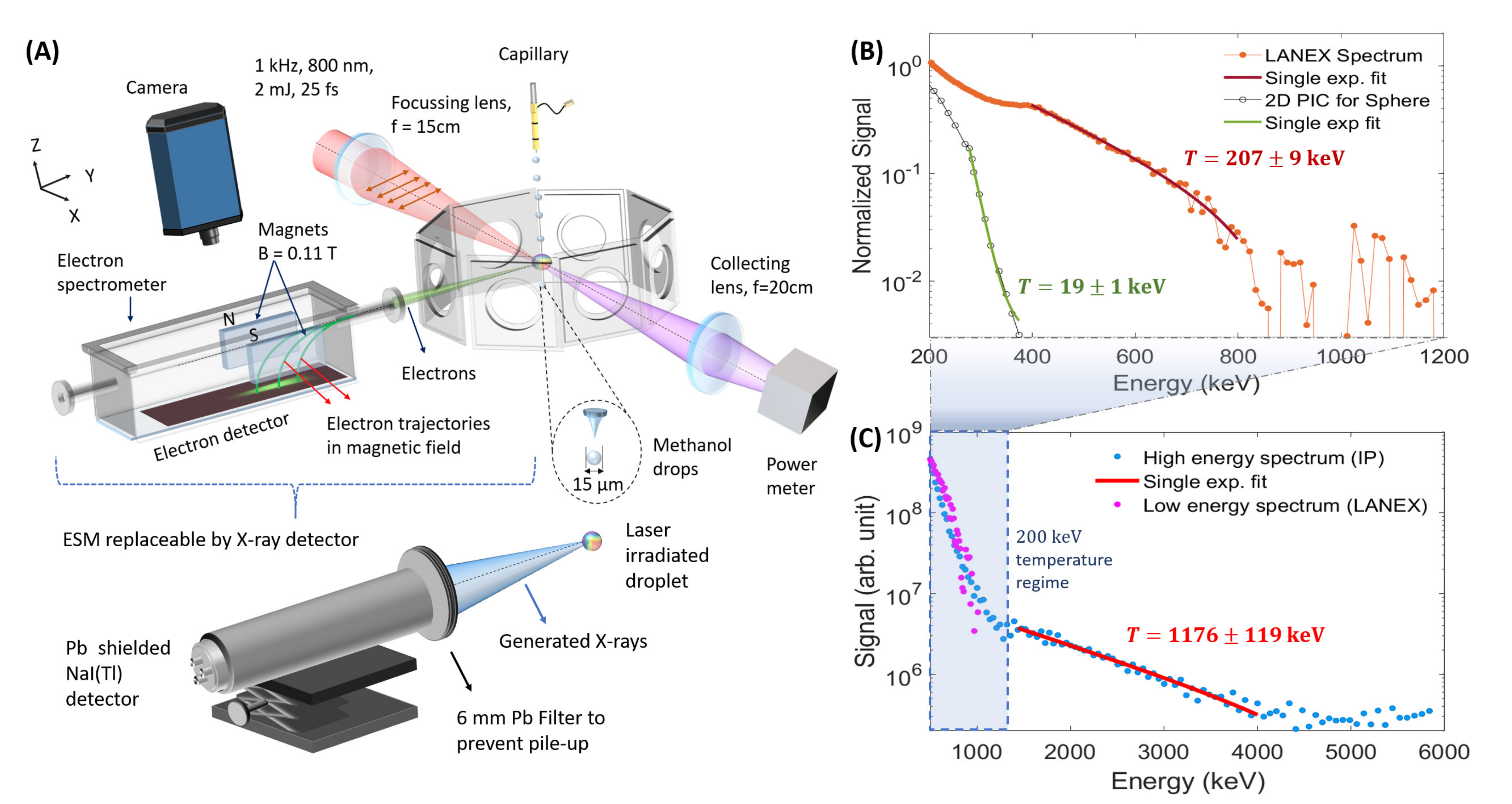}
\caption{(A) Experimental set-up. Hot electron generation is achieved by irradiating Methanol droplets at a kHz repetition rate with 800 nm, 2 mJ, 25 fs laser pulses focussed to an intensity of $4 \times 10^{16}$ W/cm$^2$. Electron and X-ray spectrum measurements are performed using a magnetic field (0.11 T) bending electron spectrometer (ESM) with LANEX or IP as detectors and a NaI(Tl) detector, respectively. (B) Electron spectrum acquired with a short acquisition (8 s) detector (LANEX) shows a 200 keV temperature component(orange circles). This is comparison to the temperature anticipated from 2D-PIC for a spherical drop(black circles) is atleast an order of magnitude higher. (C) Additionally, measurement of the same electron spectrum for 900 s with a high dynamic range detector(IP), also reveal the presence of a relativistic temperature component larger than 1 MeV. The LANEX spectrum from 1(B) (magenta circles) is superimposed on the IP spectrum (blue circles) to highlight the 200 keV temperature component and emphasize the the lower dynamic range of the LANEX detector.}
\end{figure}

\twocolumngrid 

\noindent Investigation of the laser-pulse parameters reveals that the uncharacteristically high T$_{hot}$ is attained only in the presence of a weak pre-pulse 4 ns ahead of the main pulse. A collinear two pulse experiment is set-up, to systematically change the pre-pulse intensity and determine the optimum pre-pulse strength for maximum X-ray/electron generation.  A clean laser beam(with no-pre-pulse) is split into two beams with a beam splitter: a pre-pulse with  10\% energy and a delayed main pulse. Then intensity and polarization of the pre-pulse is controlled using a combination of a polariser and a half-wave plate while the time difference between the pre-pulse and the main pulse is varied using a delay-stage. The two beams are later combined by another beam splitter to maintain the collinear geometry. Fig. 2(B) shows the variation in X-ray yield as a function of pre-pulse percentage, for a main pulse of energy 1 mJ incident 4 ns after the pre-pulse. It is observed that the X-ray emission is negligible in the absence of any pre-pulse but rises sharply and saturates when the pre-pulse is about 4-5$\%$ of the main pulse.  For all other results, we therefore use a pre-pulse containing 5$\%$ of the main pulse intensity. The incident pre-pulse is expected to increase main-pulse absorption through pre-plasma generation\cite{tg1, anand1}. However, 2D PIC simulations (Fig. S5) reveal that the presence of the pre-plasma alone, fails to generate relativistic electron temperatures at the experimental intensities. The peak plasma density and laser intensity is varied over a large range, to show that even at a laser intensity of about $\sim 1.5 \times 10^{17}$ W/cm$^2$, the electron energies are at least ten times smaller than those observed in our experiments.\\

\onecolumngrid

\begin{figure}[h]
\includegraphics[width=\columnwidth]{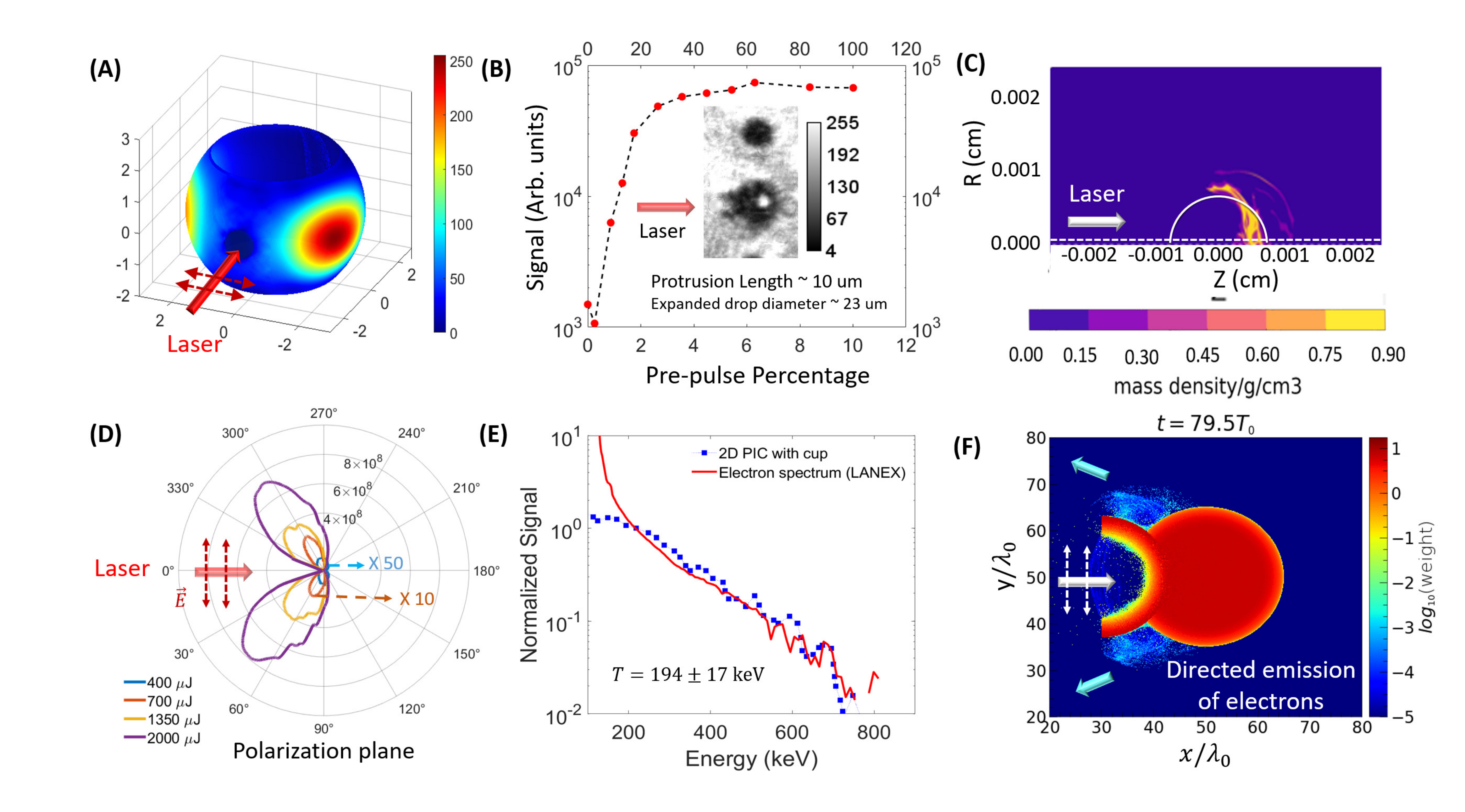}
\caption{(A) A 2D map of the experimentally obtained angular distribution of the emitted electrons. The incident laser and the polarization direction are demarcated with solid red arrow and broken red arrows respectively. The origin (0,0,0) denotes the position of the irradiated droplet. The electrons are emitted primarily in the polarization plane a oriented long the backward direction at $\pm$ 50$^\circ$ w.r.t. to laser incidence. (B) The effect of pre-pulse percentage on the X-ray yield. The corresponding pre-pulse energy in $\mu$J is indicated on the top axis. Inset shows the shadowgraph of the droplet that would be seen by the main pulse. First drop shows the shadow of typical droplet without the pre-pulse irradiation(more details given in supplement). (C) h2d radiation-hydrodynamic simulation  shows a concave surface formed 4 ns after the pre-pulse interaction.  The pre-pulse simulated by a main-pulse replica having 5$\%$ of the main-pulse intensity. The white semi-circle indicates the initial droplet geometry  and position before pre-pulse interaction. The white dotted line demarcates the axis of symmetry. (D) Variation of electron emission angle as a function of laser intensity. The yields at 0.8$\times$ 10$^{16}$ W/cm$^2$ and 1.4$\times$ 10$^{16}$ W/cm$^2$ are multiplied by factors of 50 and 10 for data representation on the same scale. (E) Comparison of the electron energy spectrum obtained from 2D PIC (blue dots) on the modified droplet geometry shown in the inset,   at 8.5 $\times$ 10$^{16}$ W/cm$^2$  with experimentally obtained electron spectrum on LANEX at 4 $\times$ 10$^{16}$ W/cm$^2$ (red solid line).   (F) Directed electron emission as observed in 2D PIC simulations. The laser is incident along the x axis(white solid arrow) with its polarization along y axis(white dashed arrows).}

\end{figure}

\twocolumngrid

\noindent However, in a micro-spherical target, a pre-pulse (25 fs pulse replica) does more than pre-plasma generation. A pre-pulse of 5\% the intensity of the main pulse, appropriately timed, can cause significant modification of the structure of the droplet prior to the arrival of the main pulse. To visualize the pre-pulse effect on the droplet structure, a shadowgraphy measurement is implemented. Similar to the pre-pulse dependent studies, the laser pulse is split into two beams in a 8:92 ratio. The weaker beam, with an intensity comparable to the original pre-pulse, is used to simulate the pre-pulse effects on the droplet. The stronger beam is frequency doubled and used to image the transverse profile of the drop. The shadowgraphy image captured using a gated 12 bit CCD camera coupled with a 400 nm bandpass filter to reject the scattered light from the first pulse. The delay between the two pulses is varied using a delay stage to capture the target modifications caused by the pre-pulse as a function of time. A droplet without the pre-pulse is also captured for reference in the same frame. The shadowgraphy image acquired at a delay of 2.6 ns shows the formation of hollow cup/cone-like concave
structure at the front of the expanding droplet(Fig. 1(B) inset); More details are in the supplement, Fig. S2).\\

\subsection{Numerical Approach}

\noindent As is evident from the experiments, the hot electrons are generated by the interplay of the pre-pulse and the main-pulse effects on the target droplet. Initially, the pre-pulse leads to a hydrodynamic evolution of the droplet over 4 ns, followed by the main pulse interaction resulting in rapid electron generation. Therefore, to investigate and identify the underlying acceleration mechanism, a 2D hydrodynamic simulations followed by 2D particle-in-cell(PIC) simulations are performed.
\subsubsection{h2d Radiation-Hydrodynamics Simulations}
\noindent The long-term pre-pulse effects on the spherical droplet are studied using 2D lagrangian radiation-hydrodynamics simulations performed using the code h2d \cite{h2d}. The simulations are run with an ideal gas equation of state, an average-atom LTE ionisation model, and a multi-group diffusion approximation for thermal radiation. The spherical target is simulated with an initial radius of 7.5 $\mu$m and an initial density of 0.792 g/cm$^2$. The laser pulse having a 800 nm wavelength, 0.1 mJ energy, 25 fs FWHM is simulated with a sech$^2$ temporal profile and a Gaussian spatial profile with a 11  $\mu$m FWHM. The final on-target intensity is $0.2 \times 10^{16}$ W/cm$^2$, 5$\%$ that of the main pulse. Conforming to the experimental pre-pulse and main pulse delay, the simulations are run for 4 ns after the arrival of the pre-pulse, at which point a snapshot of the mass density of the mesh is taken.\\

\noindent  As expected, the hydrodynamic simulations also reveal the formation of a concave cup-like structure is also observed in the experiments. Fig. 2(C) shows the mass-density distortion, acquired from h2d simulations, 4 ns after the pre-pulse interaction. The position of the original droplet is shown by a white dotted semi-circle while the position of line-out profiles is demarcated by the white dashed lines.  Besides, target deformation, a plasma gradient is also seen to be generated by the pre-pulse (Fig. S3). Having extracted the the modified target geometry and the plasma parameters from the h2d simulations, 2D-PIC  simulations are done to explore modifications in the main laser plasma interaction process.\\
\subsubsection{SMILEI and EPOCH Simulations}
\noindent The main-pulse interaction with the pre-pulse modified droplet is studied using two PIC codes: SMILEI\cite{PIC} and EPOCH\cite{EPOCH}. Typically, a simulation box size of $100\times100\lambda_0^2$ is divided into $3000\times3000$ cells., where the droplet is placed at the center. The droplet is modelled as a pre-ionised plasma represented by 49 macro-particles per cell. Ions are treated as a stationary background. For the simple spherical droplet discussed earlier (Fig. S4(A) nocup), the plasma electron density is initialised at $n_e/n_c=10$ for $r/\lambda_0<5$ and $n_e/n_c=10\cos(\pi(r-5)/10)$ for $5<r/\lambda_0<10$, where $r=\sqrt{(x-50)^2+(y-50)^2}$ is the radius measured to the center of the droplet at $x=50\lambda_0$ and $y=50\lambda_0$. For the cup-like pre-plasma(Fig. S4(A) cup-hs3), the droplet electron density is given by $n_e/n_c=7$ for $r/\lambda_0<10$ and $r_c/\lambda_0>13$, and $n_e/n_c=7\cos(\pi(r-10)/10)$ for $r/\lambda_0<15$ and $r_c/\lambda_0>13$, where $r_c=\sqrt{(x-30)^2+(y-50)^2}$ is the radius measured to the cup center at $x=30\lambda_0$ and $y=50\lambda_0$; the cup electron density is given by $n_e/n_c=7\exp(-(13-r_c)^2/4)+0.01$ for $x/\lambda_0>30$ and $r_c/\lambda_0<13$. A laser of normalized vector potential $a=a_0\sin(\pi t/\tau)^2$ is launched from the left boundary, where $a_0=0.2$ and  $\tau=25T_0=25\lambda_0/c$  correspond to a peak intensity of $8.56\times 10^{16}\rm$ W/cm$^2$ and a pulse duration of 24.3 fs (FWHM in intensity profile), respectively. The laser pulse is modelled by a Gaussian transverse profile with a waist radius of $W_0=12\lambda_0$. A virtual circular detector/screen surrounding the target is used to record the energy and angular distribution of the ejected electrons. The laser parameters in EPOCH simulation are kept identical to the SMILEI simulation. In addition in the EPOCH simulations, an ionization module is included by the addition of a background of Ar(has similar ionization energy to N) neutral atoms, represented by 49 macro-particles per cell in the simulations corresponding to a gas density of  10$^{17}$gm/cm$^3$. Both simulations provide similar results as shown in Fig. S10.\\
\noindent PIC results with the modified target, presented in Fig. 2(E), show a dramatic rise in the electron temperature from 19 keV (seen with the spherical undistorted target Fig. 1(B)) to 194 keV along with the characteristic directed emission(Fig. 2(F)). The compelling correlation with the experimental spectrum in the 200 keV-1 MeV region and the electron angular distribution validates the decisive roles played by the plasma-filled cup in the electron. A range of other target geometries and plasma density profiles that are explored, in the supplement, to determine both the optimal conditions and the sensitivity of the results to modification of the target(see Supplement Fig. S4).
\section*{Discussions}
\noindent Having established the optimal target conditions required for the hot electron generation, the underlying interaction is now explored. A one-to-one correlation, presented below, of the experimental and simulation results with the prevalent properties of parametric instabilities, identify TPD to be the dominant acceleration mechanism.\\
\noindent i) In TPD,  a laser light of frequency $\omega$ incident on an underdense plasma decays via the excitation of two plasmons of frequency $\omega/2$ at the quarter-critical( $n_c/4$) plasma density layer , where n$_c$ $\sim$ 10$^{21}$/$\lambda_{\mu m}$ cm$^{-3}$ for the laser wavelength, $\lambda_{\mu m}$. The presence of a pre-plasma gradient with a high scale-length enhances the non-linear growth of the generated plasmons, following which a plasmon decay releases high energy electrons.  This necessity of a long scale-length pre-plasma for hot electron generation, is also revealed in both the experimental (Fig.  2(B), Fig. S6) and numerical (Fig. S4 (B)) studies of the laser-droplet interaction.\\
\noindent ii) The symmetry of the systems allows the generation of directed plasmons along the 45$^o$ direction on either sides of the laser propagation axis \cite{TPD, angle}. A spatio-temporal analysis of the electric fields from the PIC simulation results, in and about the droplet cavitation surface, also reveals the same. The forward moving plasma waves  are shown to impinge on the cup boundary releasing hot electrons(Fig. S7). In comparison, due to the absence of a critical density barrier, the backward directed plasma waves are shown to grow unhindered resulting in resonant excitation of electrons to relativistic energies. For accelerating electrons by this mechanism to a sub-relativistic(200 keV) and a relativistic(511 keV) temperature component, plasma waves with wave vector amplitudes(k) of 4.5$\times$ 10$^{6}$ m$^{-1}$ and $5.64 \times 10^6 m^{-1}$, respectively, are required. Fourier analysis of the plasma wave dispersion at the apex of the cup(Fig. 3(F)), reveal wave vectors with the appropriate magnitudes to be maximised along the $\pm$ 45$^\circ$ and $\pm$ 135$^\circ$ directions in the laser polarization plane.\\

\onecolumngrid

\begin{figure}[htbp]
\centering
\includegraphics[width=\columnwidth]{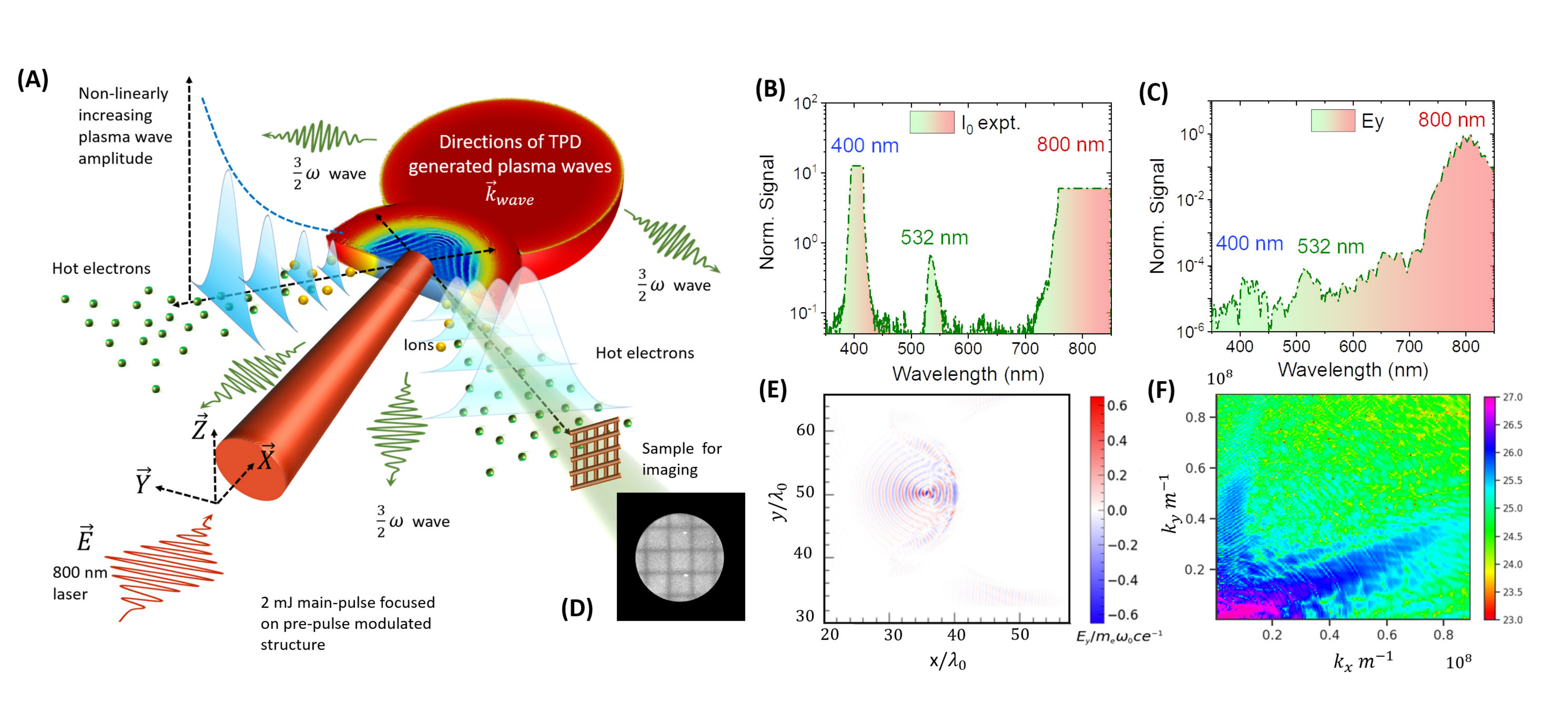}
\caption{(A) Schematic of hot electron generation mechanism in the pre-pulse modified droplet. The incident laser pulse decays via TPD in the pre-plasma gradient of the cup at n$_{cr}/4$. The plasmons generated(marked with black dashed arrows) resonantly oscillate the electrons in the plasma. As they are damped, the accelerated electrons are ejected as backward directed jets,  along with 3/2 harmonic emission. (B) The experimentally measured optical spectrum along the backward-propagation direction(opp. to normal incidence). I$_0$ denotes the normalized incident intensity on the spectrometer. (C) Fourier spectrum of the electric field components obtained along of the backward-propagation direction. As the 3/2 harmonic light propagates primarily along the negative -x axis in this region, the 532 nm peak is distinct in the E$_y$ component as opposed to the E$_x$ component of the wavelength spectrum. (D) The backward directed electrons at $\pm 50^o$ can be used for electron radiography. This is demonstrated with and electron image of a 41 $\mu$m thick Ni wire grid acquired with a 100 ms LANEX exposure. (E) From 2D PIC, the back-reflected light inside the cavity is shown to undergo a 10-fold  intensity enhancement at the apex of  the concave surface cavitation . (F) 2D k-space spectrum of the plasma fields in the cavitation region (demarcated by white box in Fig. S8(inset)) shows generation of directed plasma waves with wave vector amplitudes capable of generating the observed relativistic electrons.}
\end{figure}
\pagebreak

\twocolumngrid 

\noindent iii) But in conventional targets, TPD, is unable to substantially enhance electron temperatures with near-threshold, ultrashort pulses. We overcome this difficulty by generating the cup-shaped dynamic structure. TPD is known to be stronger for oblique incidence, where the instability interacts with longer effective pre-plasma length scales as opposed to normal incidence\cite{oblique1, oblique2, oblique3}. The cup geometry significantly lowers the TPD enhancement threshold by ensuring oblique angle incidence. This also explains the varying electron emission angle with increasing laser intensity (Fig. 2(D)). At low intensities (8 $\times$ 10$^{15}$ W/cm$^{2}$), the absence of a strong critical density concave surface initiates TPD only at high incidence angles and electrons are emitted dominantly at $\pm$ 90$^\circ$. Increasing the laser intensity, progressively drives 'perpendicular' TPD at the back of the cup, changing the emission angle to $\pm$ 45 $^\circ$ for 4 $\times$ 10$^{16}$ W/cm$^{2}$ .\\
\noindent iv) In addition to driving oblique incidence TPD, the high-density cup-structure, also reflects the incident laser pulse and focuses it to ten times the incident intensity (Fig. 3(E)). Subsequently, the phase-space plots (Fig. S and S9) of fields show the reflected laser pulse boosting the backward propagating plasmons, enabling the generation of the 1 MeV temperature component. However, since the enhanced intensity is confined, both temporally and spatially, to a small volume fraction, the 1 MeV temperature component, is not captured within the dynamic range (2 orders of magnitude) of the 2D PIC electron spectrum; it is however evident from the 2D k-space map (Fig. 3(F)) of the laser-plasma interaction.\\
v) Lastly, the TPD induced plasma waves are expected to interact with the incident light and stimulate Raman scattering. The $\omega/2$ plasma wave coupling with the incident light at $\omega$ (800 nm) should generate $3/2 \omega$ light at about 530 nm\cite{532nm}. Figs. 3 (B) and (C), show the 3/2 $\omega$ emission observed both in experiment and simulations  respectively, providing direct evidence of the TPD mechanism. Both the experimental and the numerical spectrum from the droplet, are captured along the back-reflected direction(opposite to the laser incidence). As the electric field component is always perpendicular to the direction of propagation, for backward propagation only E$_y$ component of $3/2$ harmonic observed from the numerically calculated Fourier spectrum. While at the side of the target both E$_x$ and E$_y$ spectrum show the existence of 532 nm peak(Fig. S1).

\onecolumngrid  
 
\begin{figure}[h]
\centering
\includegraphics[width=\columnwidth]{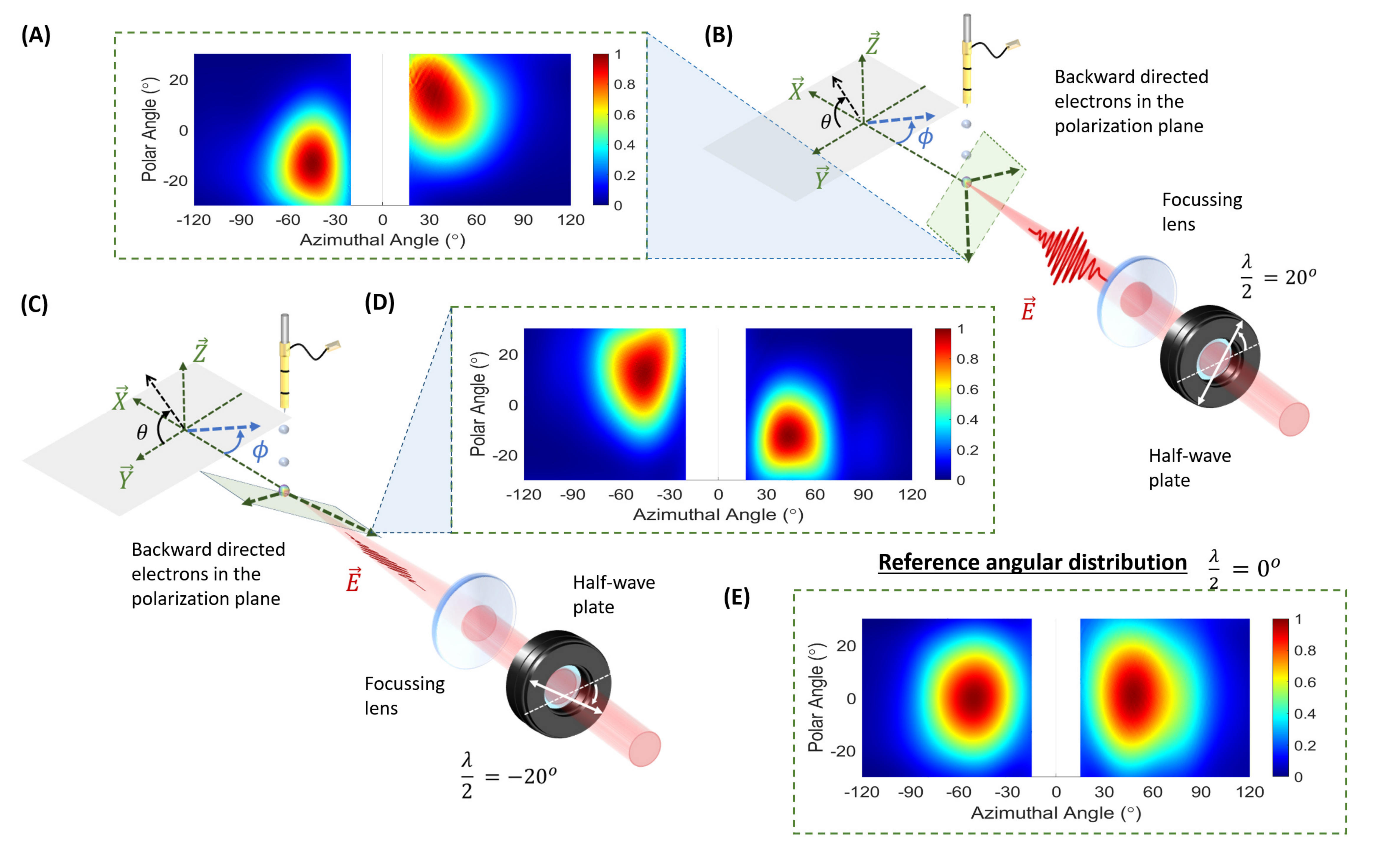}
\caption{(A) Angular distribution when the laser polarization is rotated 40$^o$ to the plane of incidence. (B) Experimental schematic showing the direction of incident laser polarization(marked also with the HWP axis) and the subsequent orientation of the electron emission plane(green rectangle).(C) Experimental schematic indicating the orientation of the electron emission plane(green rectangle) when the incident electric field rotated in the opposite direction(40$^o$) with respect to the plane of incidence.(D) Angular distribution when the laser polarization is -40$^o$ to the plane of incidence.(E) Reference angular distribution of electrons for the laser polarization in the plane of incidence.}
\end{figure}
 
 \pagebreak 
 
\twocolumngrid

\noindent However, as the two plasmons in TPD are generated in the plane of laser polarisation, a final experimental verification of the mechanism would be to rotate the incident laser polarization and observe the change in the electron emission plane.   Since, there is negligible backward and forward directed electrons (Fig. 2(A)), Fig. 4(A) and Fig. 4(D) measurements have be performed with the IPs extending from the azimuthal angle of 20$^\circ$ to 120$^\circ$ on either side of the droplet. Fig. 4(A) and Fig. 4(D) show the angular distributions when the incident polarization is rotated first to 40$^\circ$ (Fig. 4(B)) and then to -40$^\circ$ with respect to the incidence plane(Fig. 4(C)). The electron angular distribution with the laser polarization in the plane of incidence is shown for reference in Fig. 4(E). The synchronous rotation of the electron emission plane with the laser polarisation provides the final verification for TPD as the dominant electron acceleration process.\\

\noindent A summary of the entire mechanism (with the actual plasma wave profiles from the 2D PIC simulations) is illustrated in Fig. 3(A). First, the spherical droplet interaction with the pre-pulse creates a concave surface and a low density plasma. The concave geometry and the pre-plasma gradient, optimises the growth of instabilities during the main pulse interaction, following which Landau damping or wave-breaking eject the observed, relativistic temperature electrons. 
The scheme very well explains why:  (a) the generation of hot electrons requires the presence of both cavitation and a plasma density gradient in the cup, (b) the electrons are emitted at $\pm$ 50$^\circ$ with respect to the incident laser direction in the plane of polarization (c) the angle of dominant electron emission changes with the laser intensity, from 90$^\circ$ at the intensity threshold to 30-60$^\circ$ at higher intensities.  Aside from the key result of enhancing the electron temperature by 2-3 orders of magnitude, we demonstrate the suitability of such size-limited droplet targets for producing high-flux, directional electron beams for radiographic applications. Fig. 3(D) shows a well resolved lens-less transmission image of a 41 µm thick Ni wire grid, generated by electrons on a LANEX scintillation screen with a 100 ms exposure time. In addition, a high-resolution X-ray radiograph of metallic(Fig. S11(A)) and encased samples(Fig. S11(D))is also performed with the droplet source. From, the simulations the pulsed nature of the emitted electrons is established. Fig. S10(C). shows a pulse duration 43 $\pm$ 2 fs and 5.2$\pm$0.2 fs for electron energy ranges of 200-300 keV and 1-1.5 MeV, respectively, making them an ideal probe for investigating ultrafast  dynamics,
\\


\section*{Conclusion and Outlook}

\noindent

\noindent We present here a dynamic, laser-controlled, target structuring scheme for achieving relativistic electron temperatures, over few-micron acceleration length at 4 $\times$ 10$^{16}$ W/cm$^2$, only. The experimental results show that this laser-plasma optimization technique, can generate two beams of electrons, having 200 keV and 1 MeV electron temperature components with the maximum electron energy extending up to about 6 MeV with mJ/pulse 20 fs lasers.  A laser-droplet with two precisely timed, collinear pulses of appropriated magnitudes, is shown to generate a concave critical density surface with an optimal plasma gradient enable electron acceleration via the TPD instability. At oblique incidence, the threshold for the TPD instability is reduced and when coupled with the focusing effect of the pre-pulse formed concave depression in the mass-limited target, the instability growth rate is sufficient to effectively generate the hot electrons observed in the experiment. The simulations verify this TPD driven electron generation mechanism and explain the dependencies of the electron properties with the laser parameters .These results, therefore, demonstrate a simple yet elegant mechanism of accelerating electrons way beyond the ponderomotive limit, leading to highly efficient hot electron production even at modest intensities. The present features of a micro-particle with a pre-pulse induced cavitation structure, filled with a low-density plasma, thus, provide optimal conditions for beam-like hot electron generation using ultrashort(25 fs), mJ energy pulses at 100 times lower intensities than previous experiments.

 
\noindent As an outlook we expand upon two important points regarding the suitability of this scheme for applications. Firstly, our droplet source operates at 1 MHz and is  synchronized with the 1 kHz  laser, which has enabled us to acquire electron radiographs even with one laser pulse. So synchronisation of the droplet-laser hit rate with the higher repetition rate system and increasing the average current is  relatively straightforward. Though currently available MHz~\cite{MHz} lasers operate as longer wavelength, lower pulse energy and longer pulse durations, we expect most of these characteristics to be beneficial for the TPD process. A few preliminary results at these anticipated pulse parameters along with a discussion on the effect of each of these parameters on the electron energies, is presented in the supplementary material (Fig. S12 (A) and (B)). Secondly, as the electrons are directed by the laser-plasma transient fields, the electron emission is also expected to ultrashort (few tens of fs) in nature,as seen in the simulation. Combining these two features would essentially result in ultrashort electron pulses having both high peak brightness and average flux, which would be invaluable for applications like diffraction, imaging, and microscopy. Our electron imaging already shows that radiographs can be acquired within a few ms even without gating the detector. The outcome would be a pulsed electron microscopy technique based around an ultrashort burst that can tremendously reduce the dosage problems in comparison to conventional, continuous TEM sources. This could also mean microscopy without the need for cryo-cooling of the sample. It is important to keep in mind that the droplet is also a hard X-ray source, capable of acquiring X-ray images  in a few seconds even with a 1 kHz laser. The demonstration of such an acceleration mechanism with a low-cost, high-repetition-rate femtosecond laser, therefore opens an exciting and new direction for the development of genuinely compact table-top accelerators that can be driven at moderate intensities by readily available low energy, high repetition rate, lasers.




\section*{Acknowledgements}
\noindent The PIC simulations were partly supported the supercomputer center ARCHER via the Plasma HEC Consortium under EPSRC (No. EP/L000237/1). MK thanks the DAE-SRC-OI award for partial funding of the project and DAE for all the support through intramural support to TIFR.

\section*{Appendix}
\subsection*{Electron, X-ray and Optical measurements}
\noindent The electron energy measurements are performed using an electron spectrometers (ESM) with LANEX and Image plates(IP) as electron detectors.  The droplet generated electrons are collimated with a 2 mm aperture before entering the ESM. A magnet of strength 0.11 T is used to bend these electrons on to the detector, the position of incidence being proportional to the electron energy. The LANEX is used for online detection of the incident electrons within the range of 200 keV - 1 MeV. Integrated counts over 8 s exposure are collected using a 8 bit CCD, with a 540 nm bandpass filter to collect the light generated from the LANEX scintillation only. For high energy trace collection, beyond 1 MeV, longer acquisition is performed using BAS-MS image plate(IP) detectors. To prevent a reduction in IP signal by ambient light, a $10$ $\mu m$ Al foil is wrapped around the IP detector. The IPs are read using GE's Amersham Typhoon image scanner.  The final electron spectra obtained from LANEX and IP are corrected for the LANEX sensitivity values~\cite{LANEX} and the PSL sensitivity and fading time characteristics,  respectively, obtained from literature~\cite{IP1, IP2}.  To corroborate the electron energy measurements, bremsstrahlung X-ray measurements are also performed. X-rays generated from the experiment are acquired using both a MINIPIX detector (for measurement of X-rays up to 300 keV) and a NaI(Tl) detector (100 keV to 6 MeV) coupled to a multi-channel analyser (MCA). The NaI(Tl) detector is triggered and gated to acquire signal only during the main pulse incidence. To avoid pile up in spectroscopic measurements for the high energy X-rays, a 6 mm Pb filter is used to reduce the count rate to 1 count per 10 pulses. This demands an acquisition time of about 900 s to generate a background subtracted X-ray spectrum up to 6 MeV(Fig. S1). Calibration of the MINIPIX detector is performed using L $\alpha$ and L $\beta$ lines of Pb having energies of $\sim$ 10 keV and $\sim$ 12 keV, respectively. While for the NaI(Tl) detector is carried out using 137 Cs, 22 Na and 133 Ba radioactive gamma sources. The filter transmission corrections are accounted for in the reported spectrum~\cite{NIST}. 
\\
\noindent In addition to the electron energy, the spatial profile of the emitted electrons are also captured with electron angular distribution measurements. IPs spanning from 0$^\circ$ to 360$^\circ$ in the azimuthal(x-y) plane and -40$^\circ$ to 40$^\circ$ in the polar(z) plane are placed  at a distance of 3.2 cm around the droplet target. The IP geometry in the Cartesian coordinate is illustrated in Fig. 2(A), where (0,0,0) indicates the target droplet position. Each IP is wrapped with 100 $\mu$m Al foil to prevent both optical light and low energy electrons from being incident on the detector. Circular openings are made both in the entry and exit direction for unhindered propagation of the laser beam. Each angular distribution measurements are acquired for an exposure of 30 s.\\
\noindent The optical spectrum of the laser-plasma emission is measured along the back-reflected direction using a spectrometer. The supercontinuum background is captured by misaligning the laser from the droplet. The TPD optical spectrum is obtained for the optimal alignment of the laser and droplet, monitored by the maximization of the electron energy on the LANEX. Fig. 3(B) shows the supercontinuum subtracted optical spectrum as obtained from the experiments.

\section*{Competing Interests}


\end{document}



\title{Laser structured micro-targets generate MeV electron temperature
 at $4 \times 10^{16}$ W/cm$^2$}





\author{Angana Mondal$^{1,2,\dagger}$,  Ratul Sabui$^{2}$, Sheroy Tata$^{1}$, R.M.G.M Trines$^3$, S.V. Rahul$^{2}$, Feiyu Li$^{4}$, Soubhik Sarkar$^{1}$, William Trickey$^{5}$, Rakesh Y. Kumar$^{2}$, Debobrata Rajak$^{2}$, John Pasley$^{5}$, Zhengming Sheng$^{4,6}$, J. Jha$^{1}$, M. Anand$^{2}$, Ram Gopal$^{2}$, A.P.L. Robinson$^3$ and M.Krishnamurthy$^{1,2}$}


\affiliation{$^1$Tata Institute of Fundamental Research,Homi Bhabha Road, Navy Nagar, Colaba, Mumbai 400005, INDIA}
\affiliation{$^2$Tata Institute of Fundamental Research, Hyderabad, Gopanpally, Serlingampalli, Telangana, INDIA, 500107}
\affiliation{$^{\dagger}$Currently at ETH Zurich, 8093, Switzerland.}
\affiliation{$^3$Central Laser Facility, Rutherford Appleton Laboratory, Chilton, Didcot OX10 0QX, UK}
\affiliation{ $^4$Department of Physics, University of Strathclyde, Glasgow G4 0NG, UK}
\affiliation{$^5$York Plasma Institute, Department of Physics, University of York, York, YO10 5DD, United Kingdom}
\affiliation{  $^6$Laboratory for Laser Plasmas and School of Physics and Astronomy, Shanghai Jiao Tong University, Shanghai 200240, China}

\maketitle


\setcounter{figure}{0} 

\onecolumngrid

\begin{figure}[htbp]
\centering
\renewcommand{\thefigure}{S\arabic{figure}}
\includegraphics[width=\textwidth]{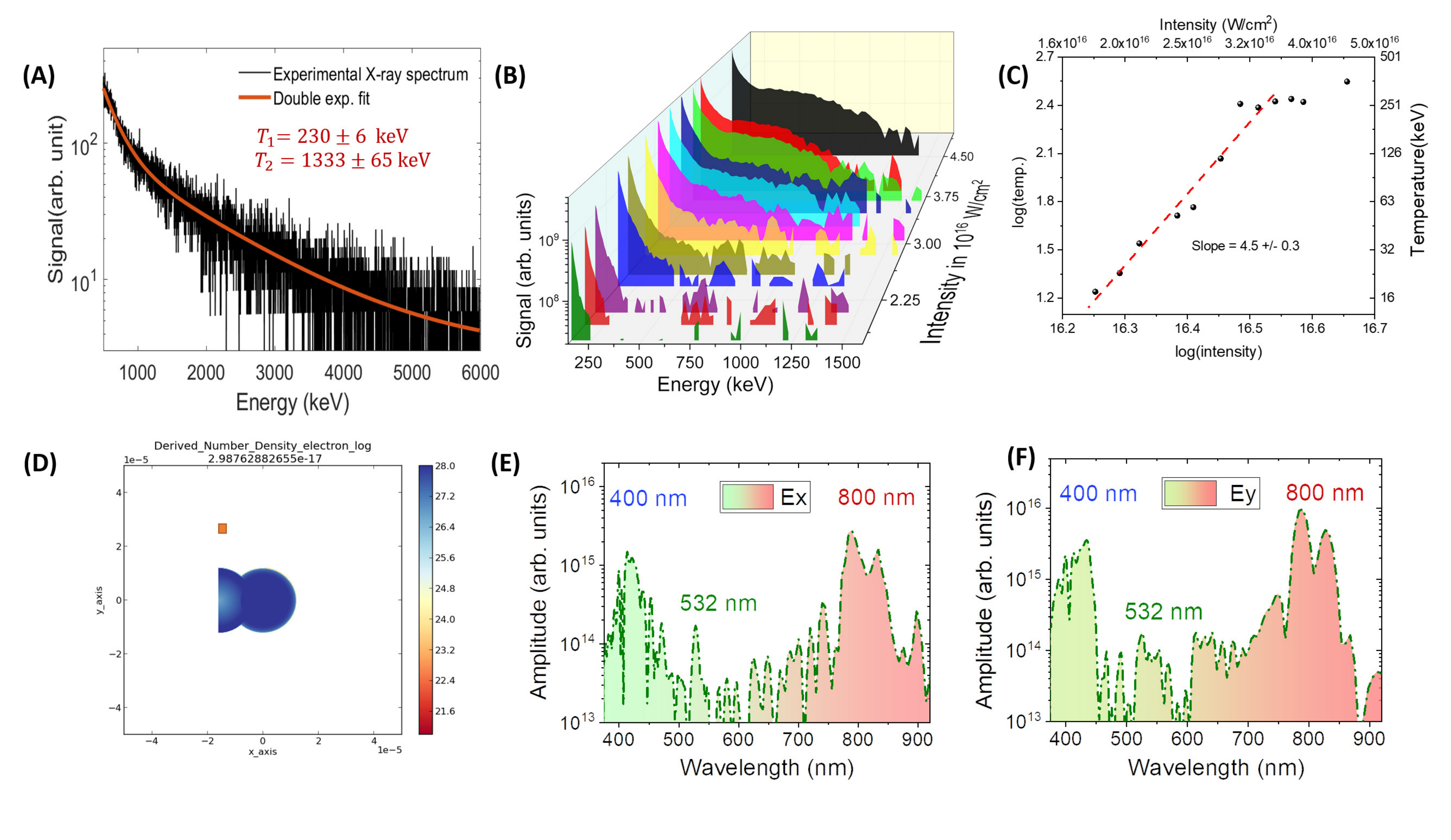}
\caption{(A) X-ray bremsstrahlung spectrum obtained using a NaI(Tl) for an acquisition duration of 900 s. A 6 mm Pb filter is used to prevent pile -up. The filter transmission has been accounted fro in the presented spectrum. As observed from the electron measurements, the X-ray spectrum also shows the existence of two hot electron components having temperature of 230 $\pm$ 6 keV and 1333 $\pm$ 65 keV, respectively. (B) and (C) The systematic increase in yield and temperature of the 200 keV temperature component as a function of laser energy. (D) Orange rectangle indicates the position w.r.t the droplet where the side-emitted 532 nm Fourier spectrum of the electric field is computed for Fig. S1(E) and S1(F).(E) The x component of the Fourier spectrum of the electric field spectrum showing the presence of 3/2 harmonic, as observed in EPOCH simulations at 90$^\circ$ from the laser incidence.(F) The y component of the Fourier spectrum of the electric field spectrum showing the presence of 3/2 harmonic, as observed in EPOCH simulations at 90$^\circ$ from the laser incidence.}
\end{figure}

\pagebreak

\twocolumngrid

\section{Supporting Experimental results}

\noindent  To verify the electron temperatures obtained from the ESM measurements (Fig. 1(B) and (C)), we also acquire the bremsstrahlung X-ray spectrum of the laser droplet interaction at 4$\times$ 10$^{16}$ W/cm$^2$ with an NaI(Tl) detector. 
Fig. S1(B) and(C) shows the variation of the electron yield and temperature of the 200 keV component as a function of the laser intensity. The electron temperature is seen to scale as I$^{4.5}$, where I is the incident laser intensity. Each spectrum is collected for an acquisition time of 8 s. Fig. S1(A) shows the background subtracted X-ray spectrum acquired over 900 s, along with the two temperature components of 230 keV and 1333 keV, respectively, similar to the electron measurements. Fig. S1(D) indicates the position w.r.t to the droplet target from where the side-emitted 532 nm spectrum is numerically observed. As opposed to Fig. 3(C), here the direction of propagation is dominantly along the y axis, therefore, the $3/2$ harmonic has a stronger E$_x$ component(Fig. S1(E)) as opposed to the E$_y$ component(Fig. S1(F)).

\section{Discussion on the shadowgraphy imaging of the droplet target modified by the pre-pulse}
\noindent In order to observe the effect of the pre-pulse on the target structure, shadowgraphy imaging is set-up. The schematic of the set-up is shown in Fig. S2(A). The principle of the measurement, in brief, is as follows: the main laser pulse is split into two beams P1 and P2. The first beam P1 containing 8 $\%$ of the beam energy and the second P2 containing the rest 92$\%$.  The beam P1 is incident first on the drop target and acts as the pre-pulse, the beam P2 then arrives at a given time delay perpendicular to P1 to image the structural change of drop due to pre-pulse. P2 is converted to 400 nm wavelength using a BBO crystal. Since the X-ray yield varies marginally with pre-pulse variation from 4 to 10$\%$ (Fig. 3(A)), we have chosen an $8 \%$ pre-pulse for our shadowgraphy experiments. The delay between P1 and P2 has been varied from 2.6 ns to 6 ns, and a CCD, time-synchronized with P2, is used to capture the droplet images as P2 arrives. A few snapshots of the droplet structure at 2.6 ns are shown in Fig. S2(B). Similar structures are also observed at later time scales. However, the shadowgraphy images do not reveal any information about the target parameters like temperature or density. For such information, we take the help of the 2D hydrocode-h2d. Also, in the main experiments, the pump-probe delay is about 4 ns and the pre-pulse intensity about 5$\%$. The shadowgraphy parameters though similar to this, is not exactly identical. We, therefore, use the hydrocode using the exact experimental parameters to find the structural modification and the associated plasma parameters of the hydrodynamically modified target.\\

\onecolumngrid

\begin{figure}[hbtp]
\centering
\renewcommand{\thefigure}{S\arabic{figure}}
\includegraphics[width=\textwidth]{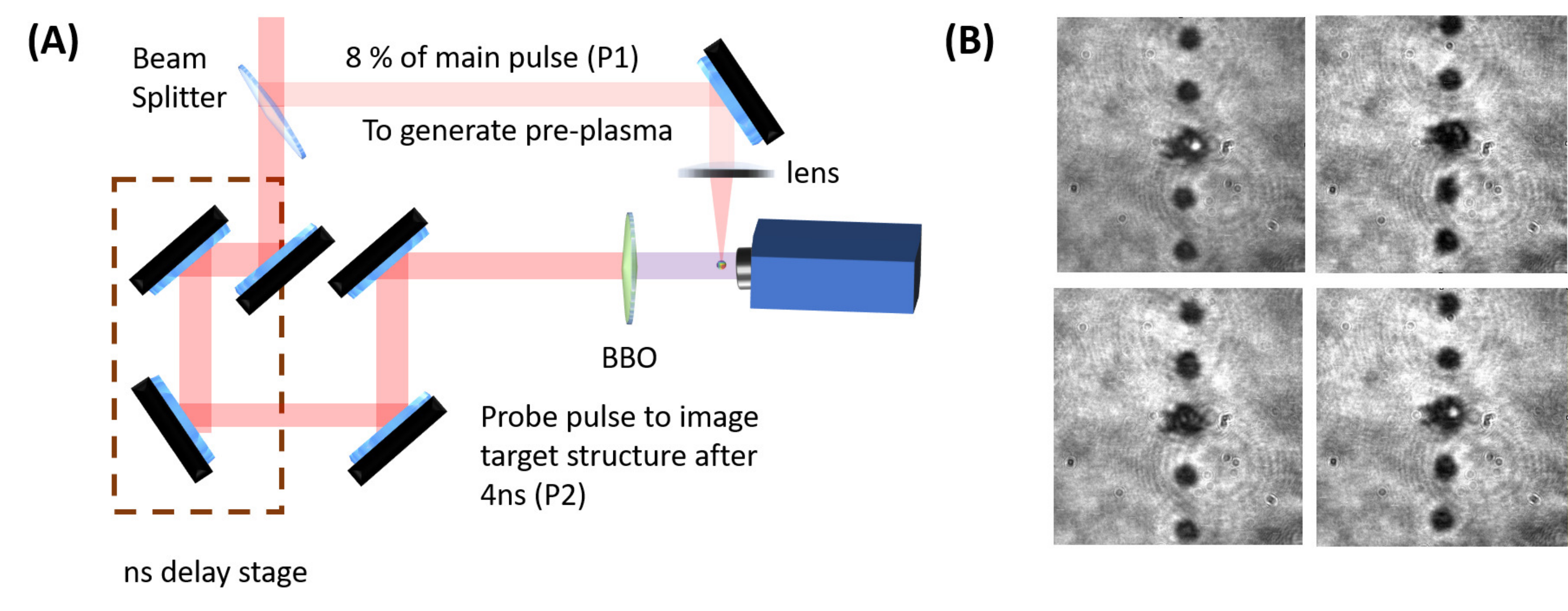}
\caption{(A) Schematic of shadowgraphy set-up. (B) Transverse shadowgraphy images (video sequence submitted in form of gif) of the droplet acquired at 2.6 ns after the pre-pulse interaction.
}
\end{figure}

\twocolumngrid

\section{Discussion on results obtained from 2D hydrocode}

\noindent The 2D radiation hydrodynamics simulations were performed with the lagrangian radiation hydrodynamics simulation code, h2d, in order to better understand the effect of the laser pre-pulse on the droplet target. The simulations were run with an average atom LTE ionisation model, and multigroup radiation diffusion. An ideal gas equation of state was employed. The laser pulse had an 800 nm wavelength, 0.1 mJ energy, and a 25 fs FWHM with a $sech^2$ temporal profile. A ray-tracing algorithm brought the energy to the target, mimicking a focusing optic employed in the experiment. The spot spatial profile was a Gaussian with an 11 $\mu m$ FWHM. The final on-target intensity was $0.2\times 10^{16}$ W/cm$^2$, 5$\%$ that of the main pulse, as in the experiment. The simulation was run for 4 ns after the arrival of the pulse, at which point snapshots of the various hydrodynamic variables were taken, since this time-point corresponds to the moment at which the main pulse arrived in the experiment. The hydrodynamic simulations also show an extended region of low density gas/plasma existing in front of the drop target(Fig. S3), which is shown to be essential for the generation of relativistic electrons.\\

\section{Discussion on effect of target structure on electron generation.}
\noindent The shadowgraphy images (Fig. S2(B)) show that the droplet is much expanded with its front surface significantly modified by the pre-pulse. In particular, some cup like pre-plasma structures with dense open walls are formed. Shadowgraphy done over several shots show that small variation in the exact positioning of the drop with respect to the laser gives shot-to-shot variation in the shape of the dynamic structures. With 2D PIC simulations, here we investigate a few possible pre-plasma modifications that resemble the shadowgraphy images and try to determine the optimized density profiles that lead to the MeV hot electrons. Fig. S4(A) shows the different configurations studied and Fig. S4(B) the resulting hot electron spectra obtained by a virtual detector surrounding the droplet. By comparing among the different cases, it is concluded that two major factors contribute to the hot electron generation, with the first being the geometrical structure and the presence of a  gradient pre-plasma. The prominence of a proper density gradient can be particularly drawn by comparing the cases cup\_ hs3 and cup\_ hs2. The latter having a constant fill-in density even gives slightly better focusing, but it produces hot electrons only about 150 keV, with the electron temperature similar to the no-cup geometry. In contrast, when filled in with a non-uniform plasma (most likely the case in experiments), the case cup\_ hs3 gives far hotter electrons with the cut-off approaching 1 MeV as seen in Fig. S5(B). The electron energy and angular distributions obtained for various laser amplitudes and maximum electron densities are summarized in Fig. S5, with the plots arranged along with the trends of increasing laser amplitude and plasma density. In general, hot electrons are more easily obtained at a higher $a_0$ and $n_{max}$. In the opposite, no electron ejection is seen for $a_0<0.3$ and $n_e<n_c$. This is because, at low densities the laser plasma coupling is insufficient to generate hot electrons. Even increasing the maximum density to $n_ {max}= 5n_c$ only generates electrons maximum electron energy of about 150 keV for $a_0 =0.1$, far below that seen in the optimized case of cup\_ hs3. To obtain hot electrons up to 1 MeV, one has to increase the laser peak intensity beyond the relativistic threshold, i.e., $10^{18}$ W/cm$^2$. \\

\onecolumngrid

\begin{figure}[htbp]
\centering
\renewcommand{\thefigure}{S\arabic{figure}}
\includegraphics[width=\textwidth]{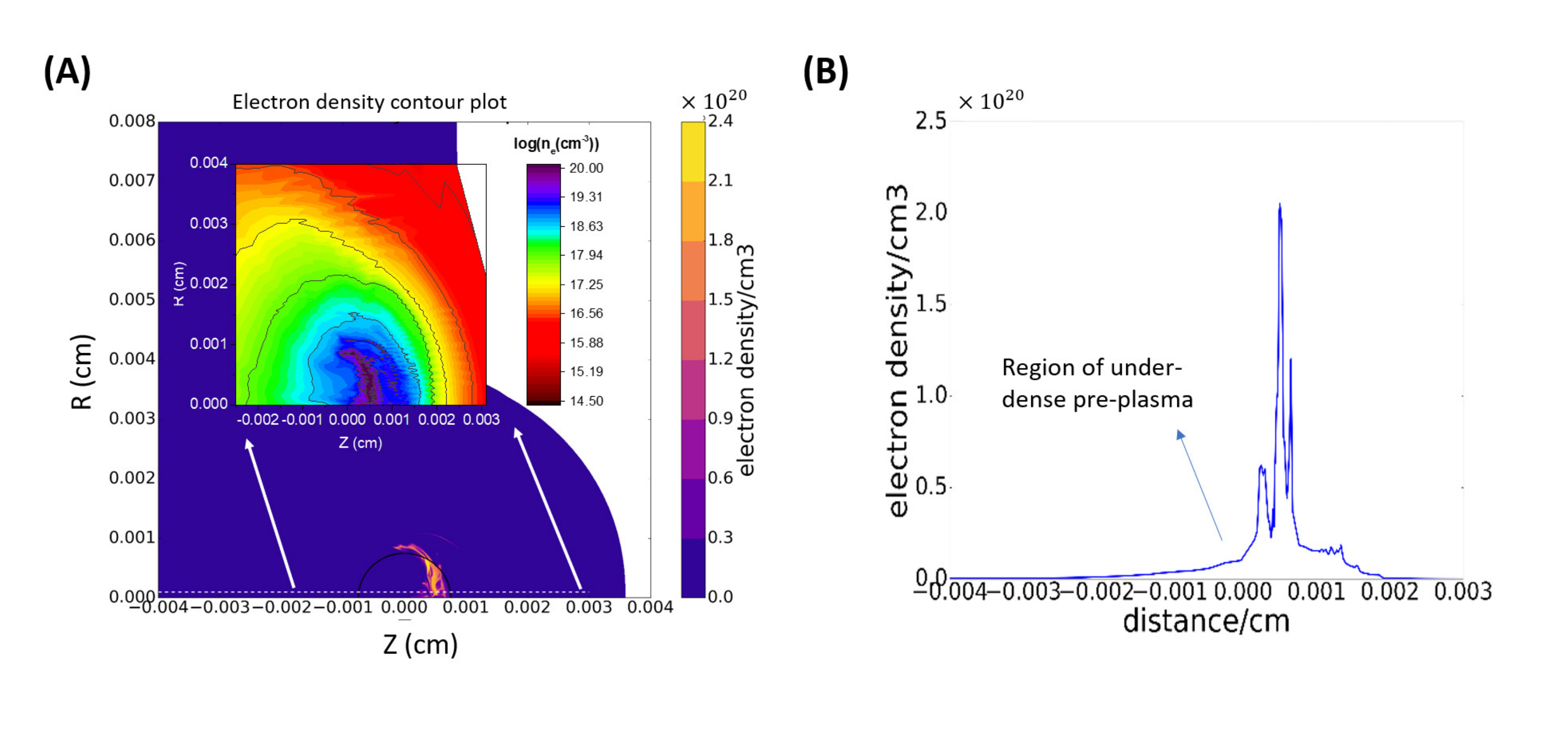}
\caption{(A) Electron density profile 4 ns after the pre-pulse interaction, as obtained from 2D hydrodynamic simulation. Inset shows the electron distribution in the arrow marked region in log scale. (B) Electron line out profile marked by the white dashed line on the 2D contour map.}
\end{figure}

\twocolumngrid

\onecolumngrid

\begin{figure}[t]
\centering
\renewcommand{\thefigure}{S\arabic{figure}}
\includegraphics[width=\textwidth]{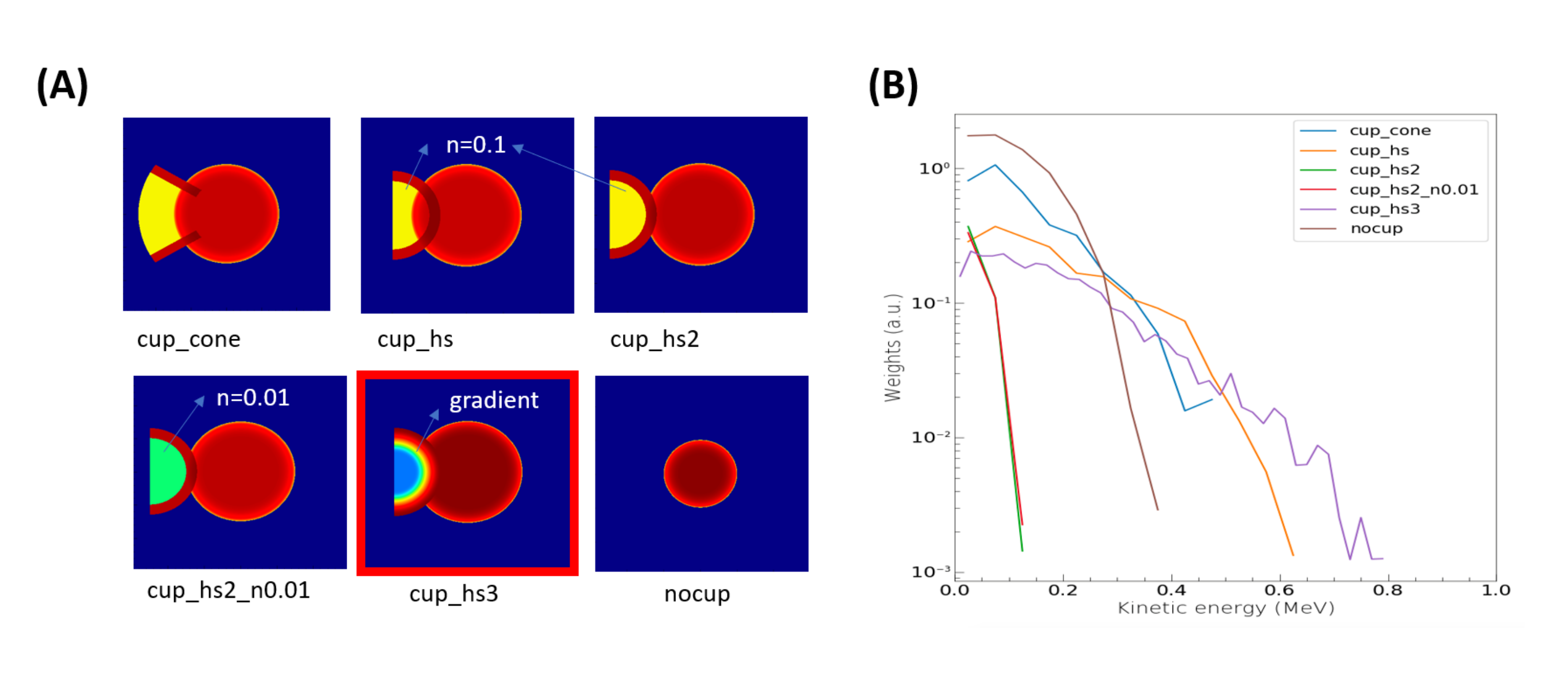}
\caption{(A) The various dynamic structures investigated by 2D PIC. The comparison suggests that the interaction with ‘high-density’ filled-in pre-plasma (preferentially having a gradient) helps produce hotter electrons. (B) Energy spectra of electrons recorded by a virtual detector for different pre-plasma conditions.
}
\end{figure}

\twocolumngrid

\onecolumngrid

\begin{figure}[h]
\centering
\renewcommand{\thefigure}{S\arabic{figure}}
\includegraphics[width=\textwidth]{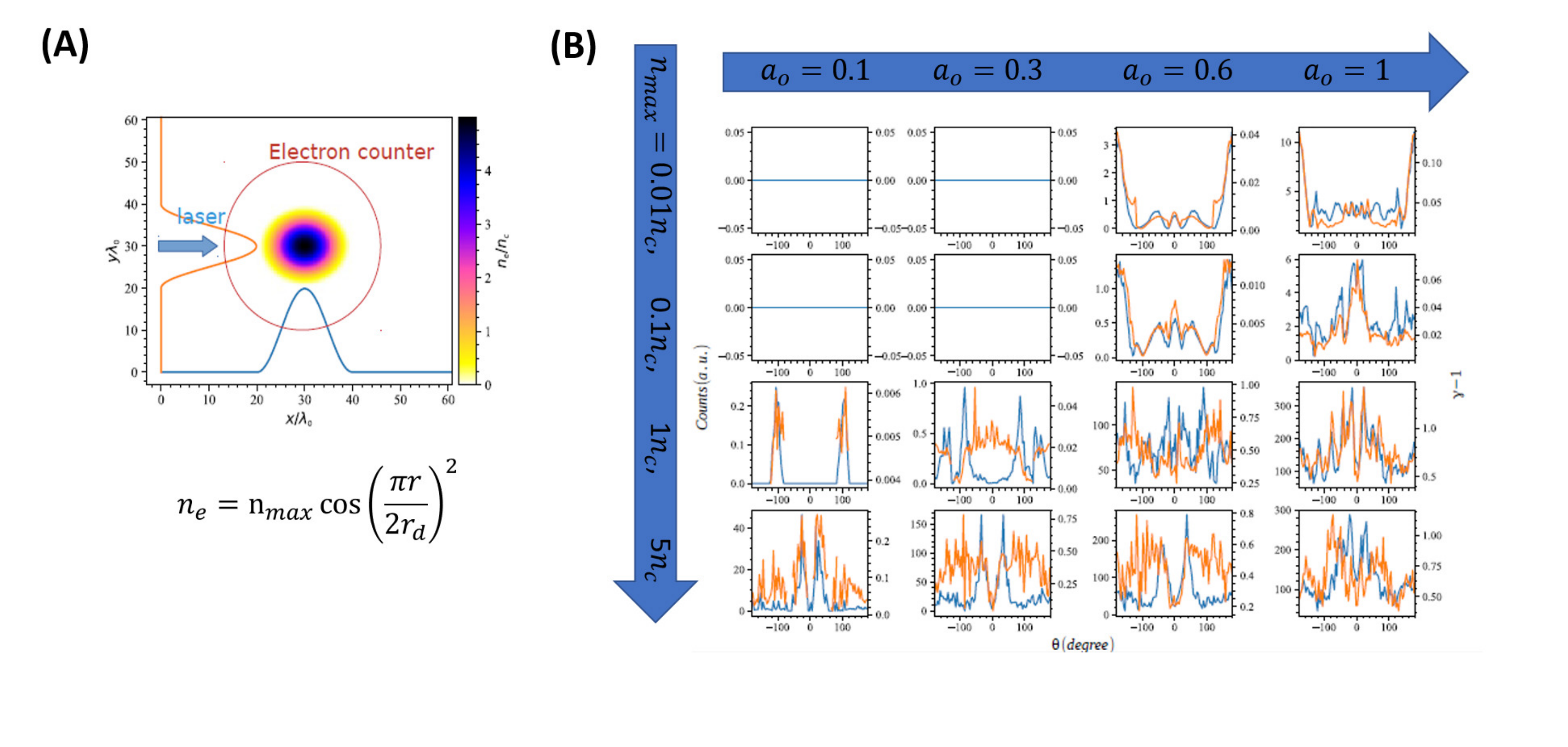}
\caption{PIC simulations done for the simple sphere at with varying density profile at different laser intensities. The left-hand side shows the initial density profile. The right-hand side shows the electron energy and angular distribution obtained at different laser intensities and initial electron densities given by $a_o$ and $n_{max}$  respectively. Where $a_o \cong 0.855 \times 10^{-9} I^{0.5}[W/cm^2] \lambda_0[ \mu m] $, is the normalized intensity, $n_{c}$  the critical density for 800 nm wavelength, r the distance from the centre of the drop and $r_d$ the droplet radius. The blue line in the plots denotes the electron counts, while the orange line depicts the energy given in terms of the rest mass energy of electrons, $\gamma -1 = 511$ keV.}
\end{figure}

\twocolumngrid

\onecolumngrid

\begin{figure}[h]
\centering
\renewcommand{\thefigure}{S\arabic{figure}}
\includegraphics[width=\textwidth]{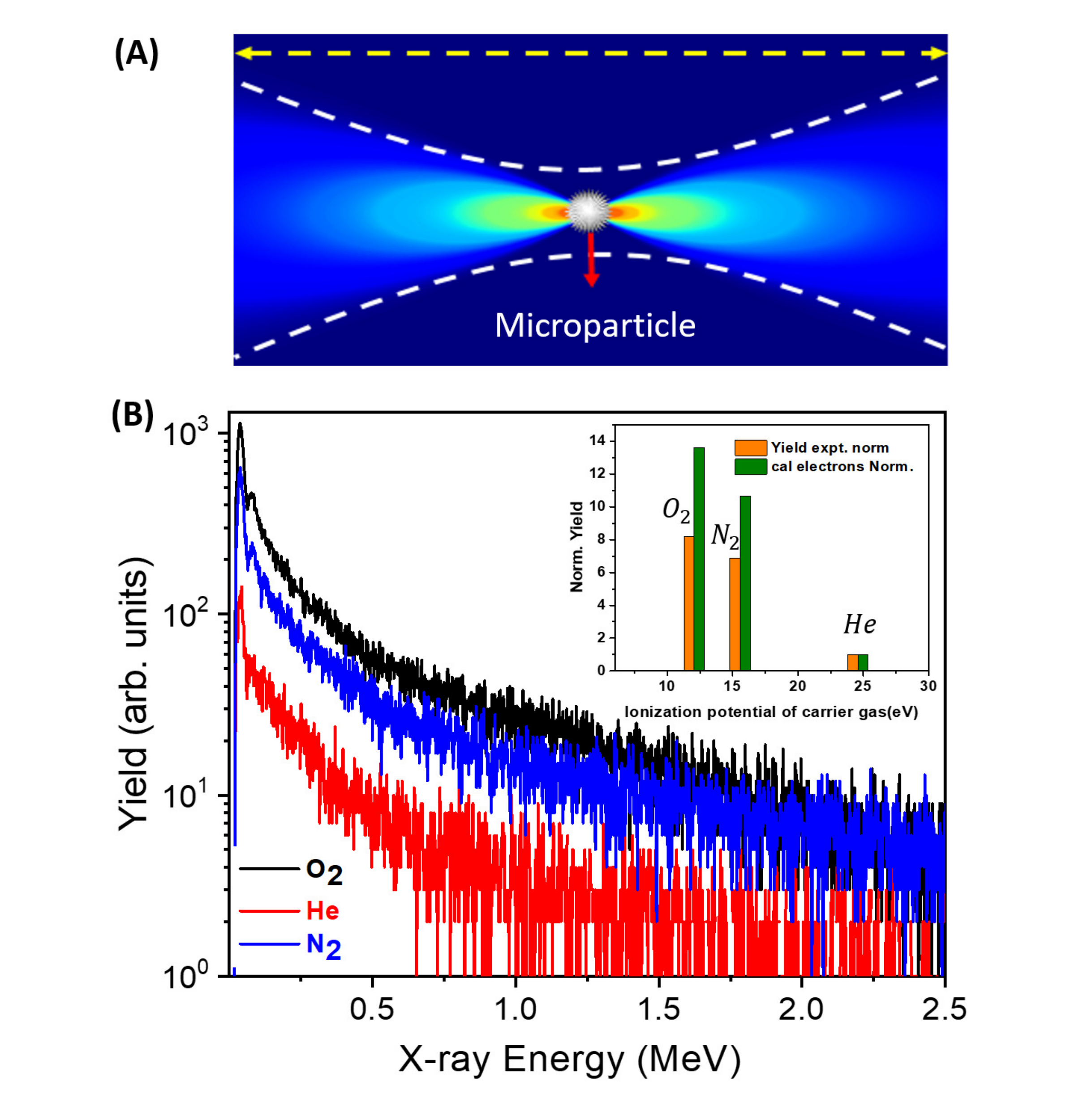}
\caption{(A) Schematic showing the laser intensity variation with the microparticle at the focal waist. The varying intensity profile generates a density gradient depending on the ionizability of the background gas. (B)Experiments done with $15$ $\mu m$ boric acid particles show that changing the ionizability of the background gas changes the X-ray yield. The inset shows a comparison of the generated X-ray yield as a function of ionization energy, with the calculated values using ADK ionization rates. The yields have been normalized to the yield values obtained for Helium.}
\end{figure}

\twocolumngrid

\onecolumngrid

\begin{figure}[h]
\centering
\renewcommand{\thefigure}{S\arabic{figure}}
\includegraphics[width=\textwidth]{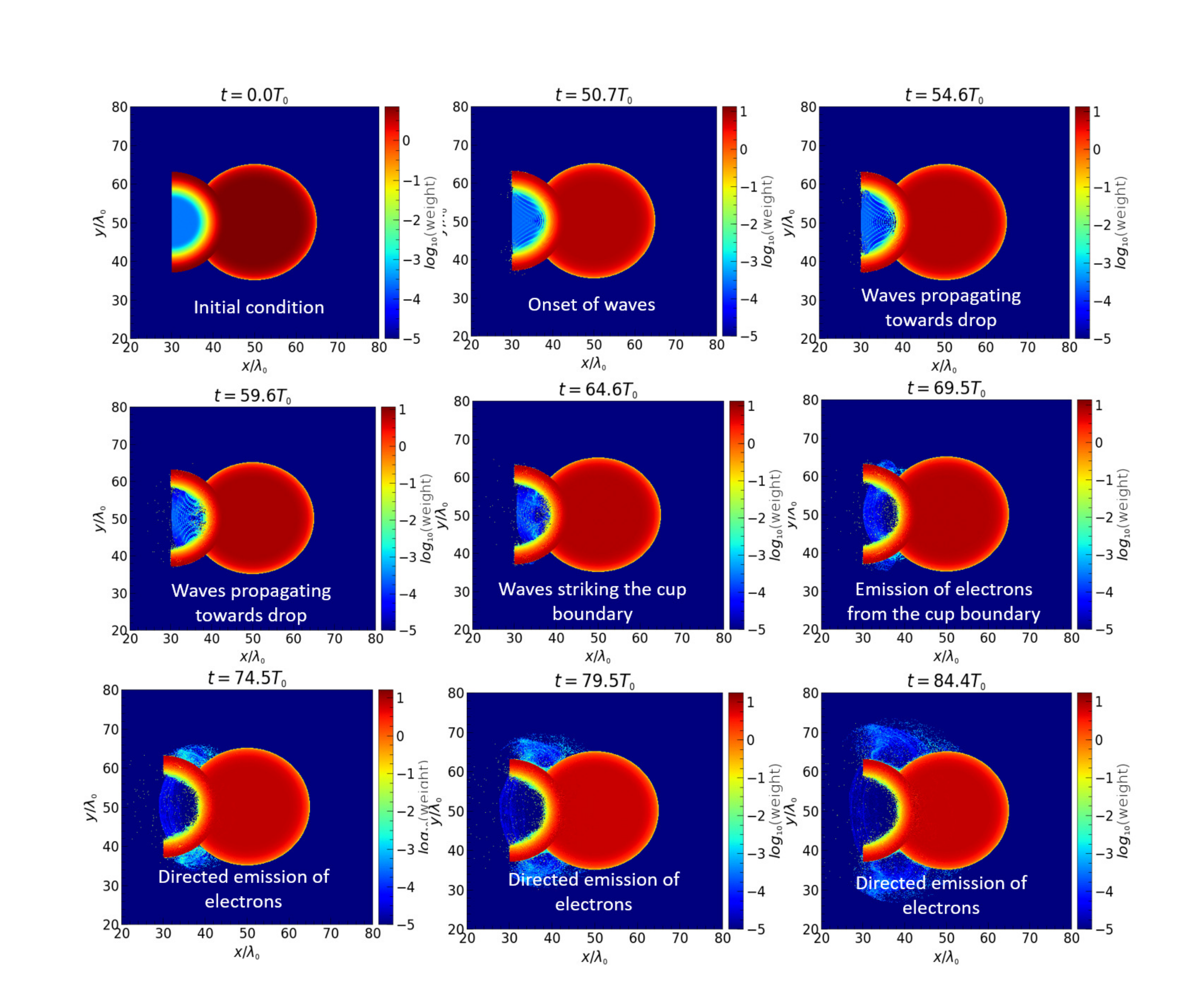}
\caption{Evolution of electron density as a function of time for $a_o =0.2$. }
\end{figure}

\twocolumngrid

\section{Role of pre-plasma}

\noindent In order to verify the importance of the pre-plasma, further experiments have been performed. Since for Methanol droplets the low density plasma surrounding the modified target is mostly dominated by the pre-pulse acting on the drop and the vaporisation of methanol, variation of this plasma density without affecting other parameters is difficult. So, to extract the sole effect of background electron density, we undertake a slightly different approach. Under similar conditions, Methanol droplets are replaced with  Boric acid particles of similar dimensions entrained in low density (0.1 Torr) carrier gas of our choice. Boric acid particles, in general, have prolate or oblate shapes. However, since the data is acquired for multiple shots and the shape and orientation of the particles to which the laser shots are incident is random, the shape effect averages out. The density gradient now depends mostly on the ionizability of the background gas. This makes it easier to change the density gradient by just changing the surrounding gas. From Fig. S6 (B), it is seen that, keeping the gas pressure constant at 0.1 Torr, as the background gas is changed from Oxygen to Helium,  the X-ray yield systematically reduces. Oxygen having greater ionizability produces a larger number of electrons as compared to Helium, which in turn increases the yield of generated X-rays. The gradient arises as the laser focal intensity varies along the laser propagation with the microparticle suspended at the focal waist. Helium with ionization potential of 24.6 eV is ionised only when the over barrier intensity threshold of $ \sim3\times 10^{13}$ W/cm$^2$  is reached, whereas Oxygen, with an ionization potential of 13.6 eV, is ionised at a much lower laser intensity of $\sim 4\times 10^{11}$ W/cm$^2$. This ionizability changes the gradient of the pre-plasma density in the cup. However, the temperatures of these spectra are identical, proving that the electron acceleration mechanism is still governed by the laser interaction with the particle itself.\\

\section{Observation of plasma waves within the cup-structure using 2D-PIC}

\noindent Fig. S7 shows, how the hot electrons are ejected in the optimized case cup\_hs3 by presenting the electron density evolution at a few selected instants. It is seen that upon initial hitting by the incident laser, plasma-waves are formed in the non-uniform plasma inside the cup. These waves propagate towards the droplet boundary and eventually crash on the hide density region, followed by the release of a copious amount of electrons. These electrons originate from the edges of the hemispherical cavitation structure and the droplet base and are beamed in the backward direction at large angles. These features are present only when the modified pre-plasma conditions are included, thus initiating oblique-incidence TPD.\\

\onecolumngrid

\begin{figure}[htbp]
\renewcommand{\thefigure}{S\arabic{figure}}
\includegraphics[width=\textwidth]{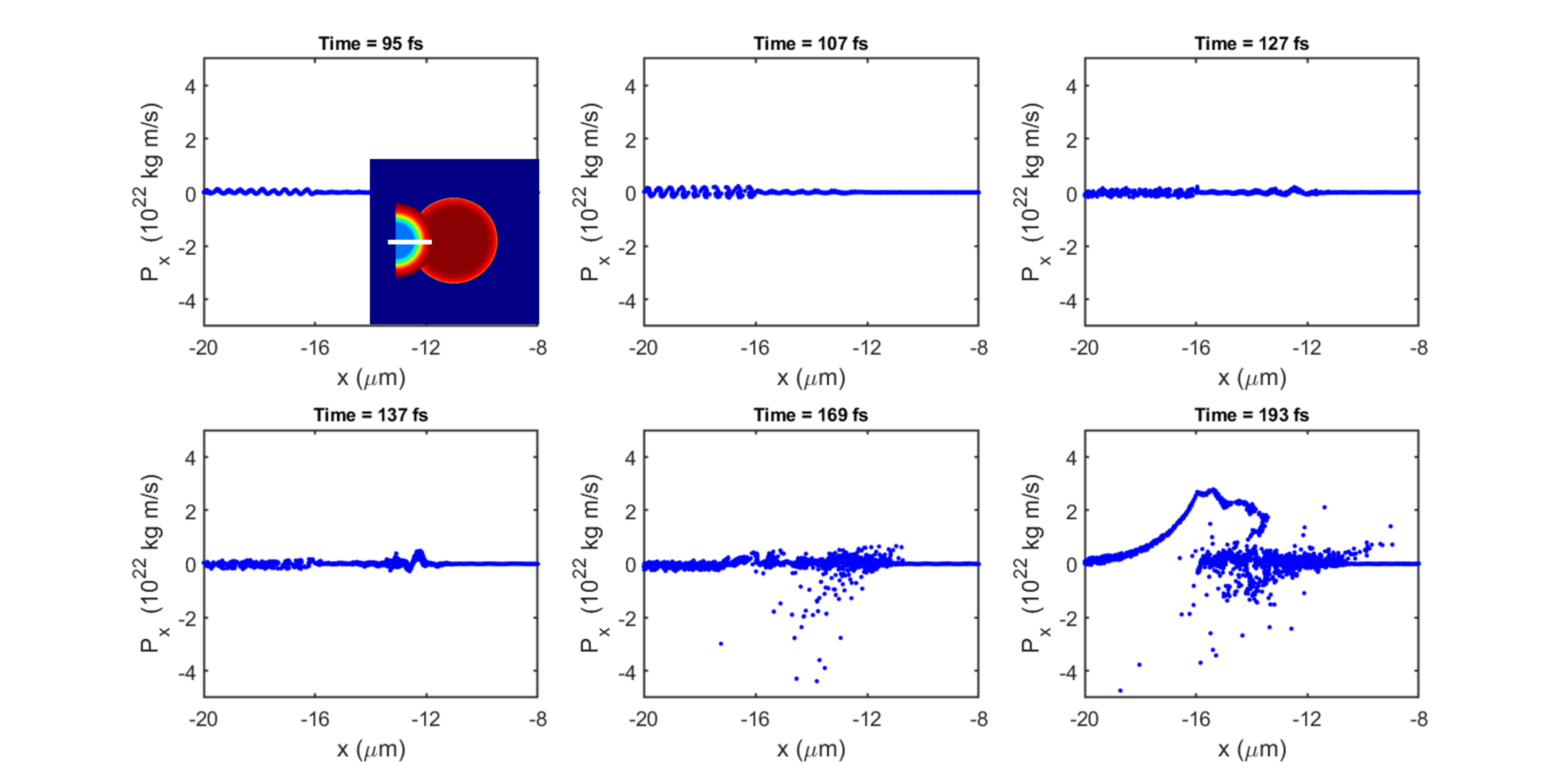}
\caption{Temporal snapshots of electron momentum(p$_x$) along the laser propagation direction as a function of x. The laser, incident from the left boundary of the simulation box is shown to excite the electrons resonantly. Inset shows the region of interest(white rectangle) considered for tracking the particle momentum as a function of time.}
\end{figure}

\twocolumngrid

\noindent This is also evident from the temporal evolution of the electron momentum components shown in Fig. S8 and S9. Fig. S8 shows the temporal snapshots of electron momentum(p$_x$) along the laser polarisation direction as a function of propagation distance x. The laser, incident from the left boundary of the simulation box is shown to excite the electrons resonantly. Inset shows the region of interest(white rectangle) considered for tracking the particle momentum as a function of time. On further propagation, the optical pulse is reflected from the overdense cup-structure.  A similar process takes place in the y direction, as depicted in Fig. S9. The reflected pulse enhances the plasma wave amplitude followed by wave-breaking or Landau damping resulting in a subsequent release of high energy electrons.\\

\section{Agreement between EPOCH and SMILEI simulations}
\noindent The phase-space plots are computed using EPOCH simulations. In addition to the particle plasma profile, the EPOCH simulations also include an ionization
module by the addition of a background of Ar(has similar ionization energy to N) neutral corresponding to a gas density of 10$^{17}$gm/cm$^3$. The agreement between the EPOCH simulation and the SMILEI PIC simulation is established from the electron temperature obtained from both these simulations as shown in Fig. S10 (A) and(B).\\

\section{Possibility of X-ray imaging}
\noindent Apart from electron imaging, that can be acquired within a minimum duration of 3 ms, we have also mentioned the possibility of X-ray imaging from the droplet generated X-rays. As a proof of principle experiment, shown in Fig. S11 (A), an X-ray-image of the $41$ $\mu m$ Ni grid has been acquired using the MINIPIX detector within the range of 5-80 keV. The optical image of the same grid taken using a microscope is presented in Fig. S11 (B). The X-ray image has been acquired at $3\times 10^{15}$ W/cm$^2$ for an exposure of 10 mins. However, at a full intensity of $4\times 10^{16}$ W/cm$^2$ such images can be acquired in about minute of exposure. Fig. S11(D) shows an X-ray radiograph within 6-9 keV of a capillary connector, acquired for an exposure of 100 s. The metallic wires  encased within the plastic connector can be clearly deciphered from the image. In addition, regions having thicker plastic casing are also captured in the radiograph as regions with greater opacity, the thickness change in the thinner and thicker plastic region being of the order of 1-2 mm.\\

\onecolumngrid

\begin{figure}[h]
\renewcommand{\thefigure}{S\arabic{figure}}
\includegraphics[width=\textwidth]{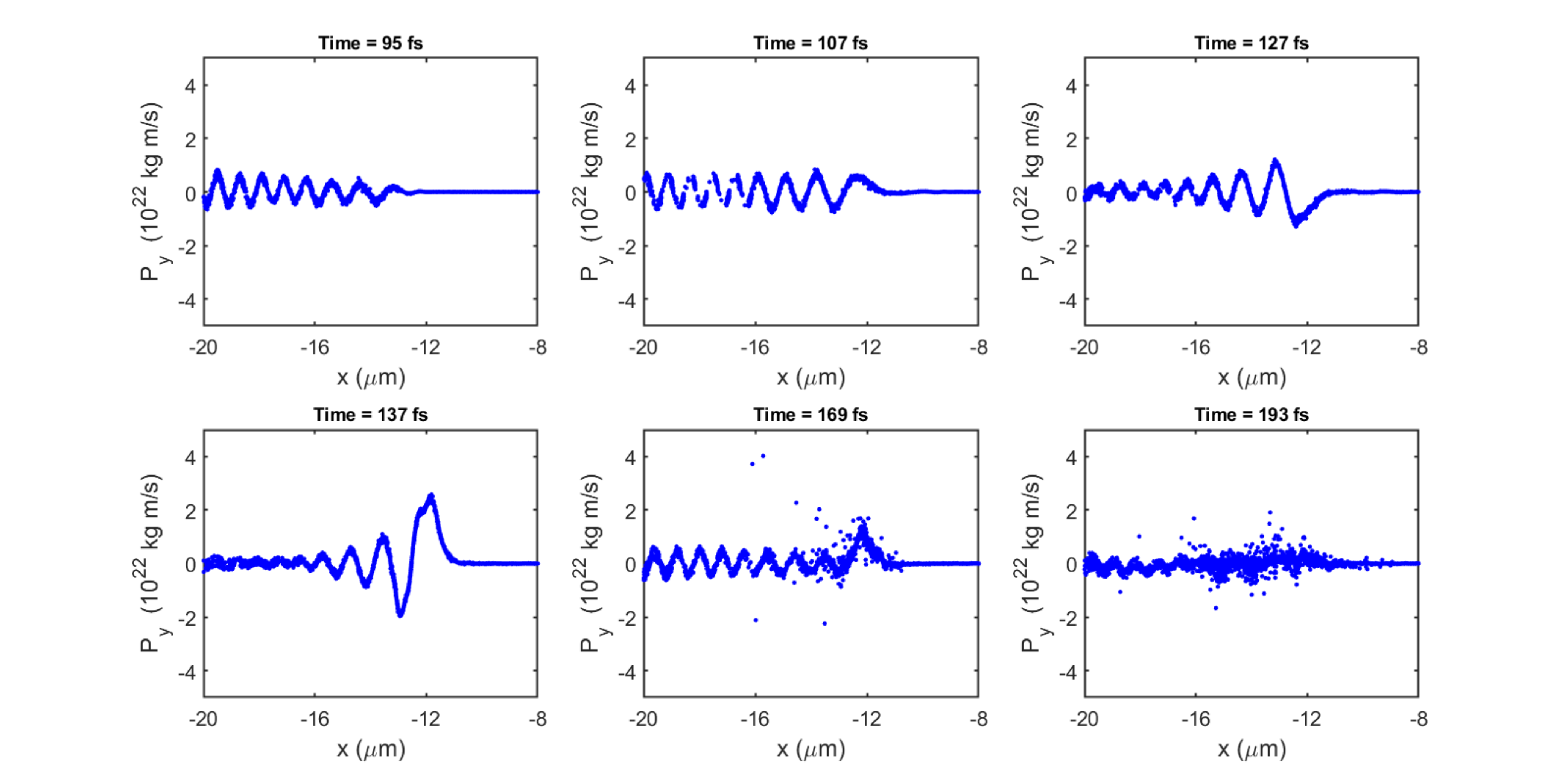}
\caption{Temporal snapshots of electron momentum(p$_y$) along the laser polarisation direction as a function of propagation distance x. The laser, incident from the left boundary of the simulation box is shown to excite the electrons resonantly. On further propagation, the optical pulse is reflected from the overdense cup-structure. The reflected pulse interacts with the oscillating electrons de-phasing them from the plasma wave resulting in high energy electron emission.}
\end{figure}

\twocolumngrid

\onecolumngrid

\begin{figure}[h]
\centering
\renewcommand{\thefigure}{S\arabic{figure}}
\includegraphics[width=\textwidth]{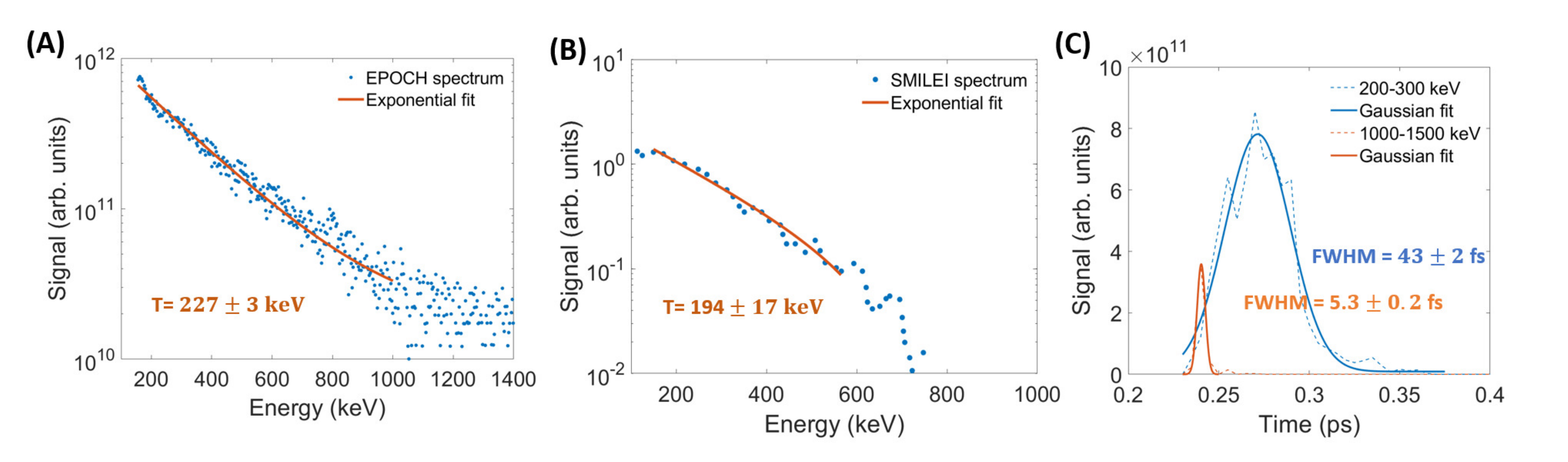}
\caption{(A)Electron spectrum obtained with 2D-PIC EPOCH simulations. (B) 2D PIC simulation electron spectrum obtained from SMILEI. (C) Temporal distribution obtained for different electron energy ranges from EPOCH simulations shows the pulsewidths varying from 40 fs for 200-300 keV electrons to 5 fs for 1-1.5 MeV electrons.}
\end{figure}
\pagebreak

\twocolumngrid

\onecolumngrid

\begin{figure}[h]
\centering
\renewcommand{\thefigure}{S\arabic{figure}}
\includegraphics[width=\textwidth]{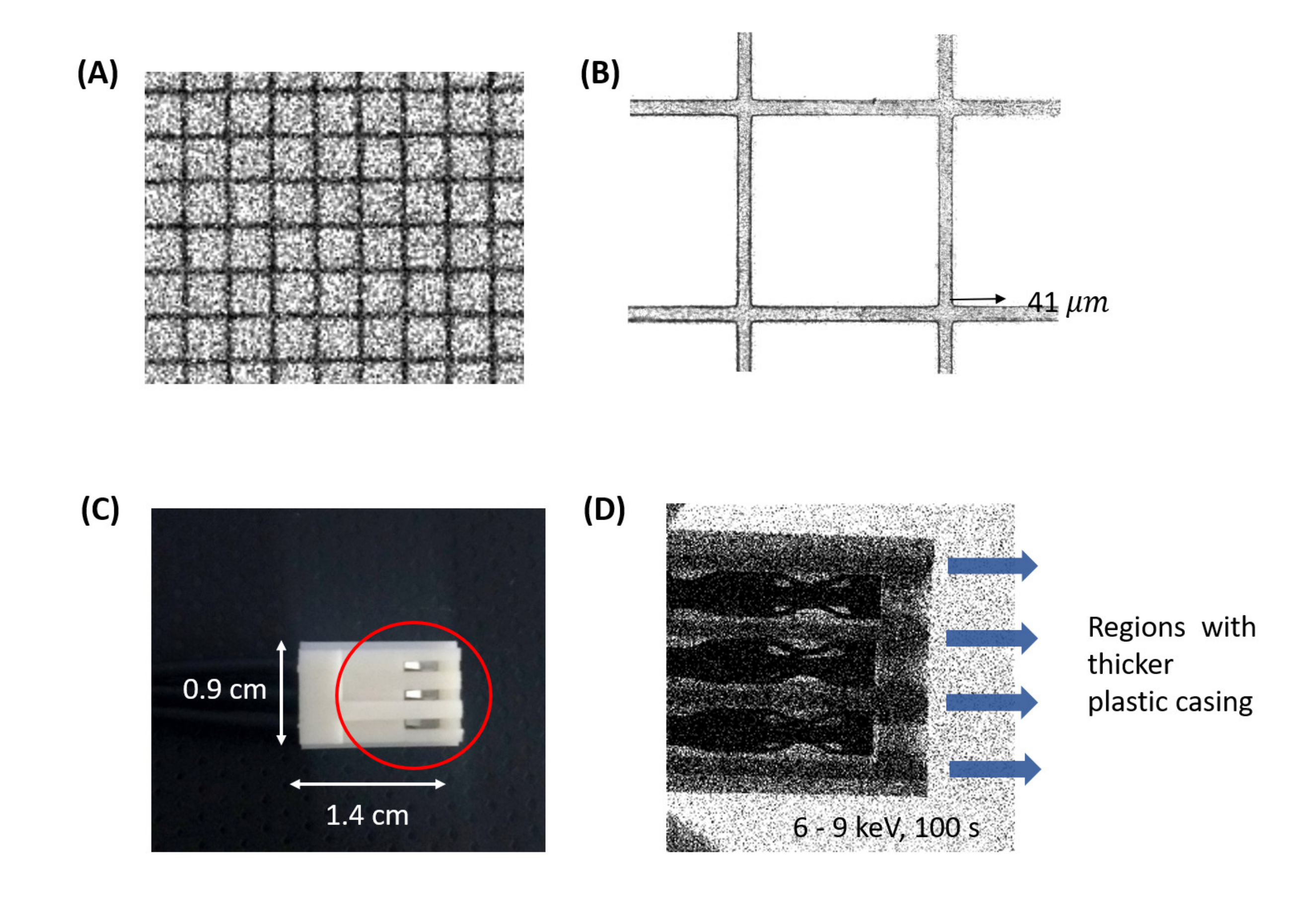}
\caption{(A) X-ray image of a $41$ $\mu m$ Ni mesh captured using MINIPIX detector within the energy range of 5-80 keV. (B) Optical image  of the same grid obtained using a high resolution microscope having 400 nm spatial resolution. (C) Capillary connector used for X-ray radiography. (D) X-ray transmission image  within 6-9 keV acquired for an exposure of 100 s using MINIPIX detector. Apart from the internally embedded metallic wires, embedded in 5 mm of plastic casing, the varying thickness of the casing is also visible in the image.}
\end{figure}

\pagebreak

\twocolumngrid

\section*{Scalability}
\noindent Our claim that the system can be extended to a MHz repetition rate is based on the frequency of the droplet generation system. Our droplet train is generated at a MHz repetition rate. Today, it is possible to purchase a laser with the following specifications (TruMicro 5080 Femto Edition: 900 fs; 200 $\mu$J/pulse Max. repetition rate of 1 MHz). Fig. S12(A) shows X-ray measurements from the droplet source observing that even at a stretched pulsewidth of 1.3 ps X-rays upto 100 keV can be generated from this source. In addtion, recent works have demonstrated lasers that operate at 260 fs; 1mJ/pulse for 1032 nm light and 1 MHz repetition rate(\textit{Optics Letters} {\bf 41}, 3439(2016)). From Fig. S12(B) it can be seen that our laser operated with
specifications close to this  with 800 nm wavelength at 1 kHz is effective in generating the electron beams of the scale presented in the manuscript.  Higher wavelengths are generally known to enhance TPD and is only expected to favour the underlying physics brought out here. So, if we have to access to these lasers we could have reported results at the 1 MHz repetition rate with no further technical issues to solve. Of course, the target deformation and the main pulse interaction will be subsequently modified because of pulse stretching, but from application perspective, our claim that the droplet plasma can be used as a MHz source of high energy X-rays/electrons is not far-fetched.\\

\onecolumngrid

\begin{figure}[h]
\centering
\renewcommand{\thefigure}{S\arabic{figure}}
\includegraphics[width=\textwidth]{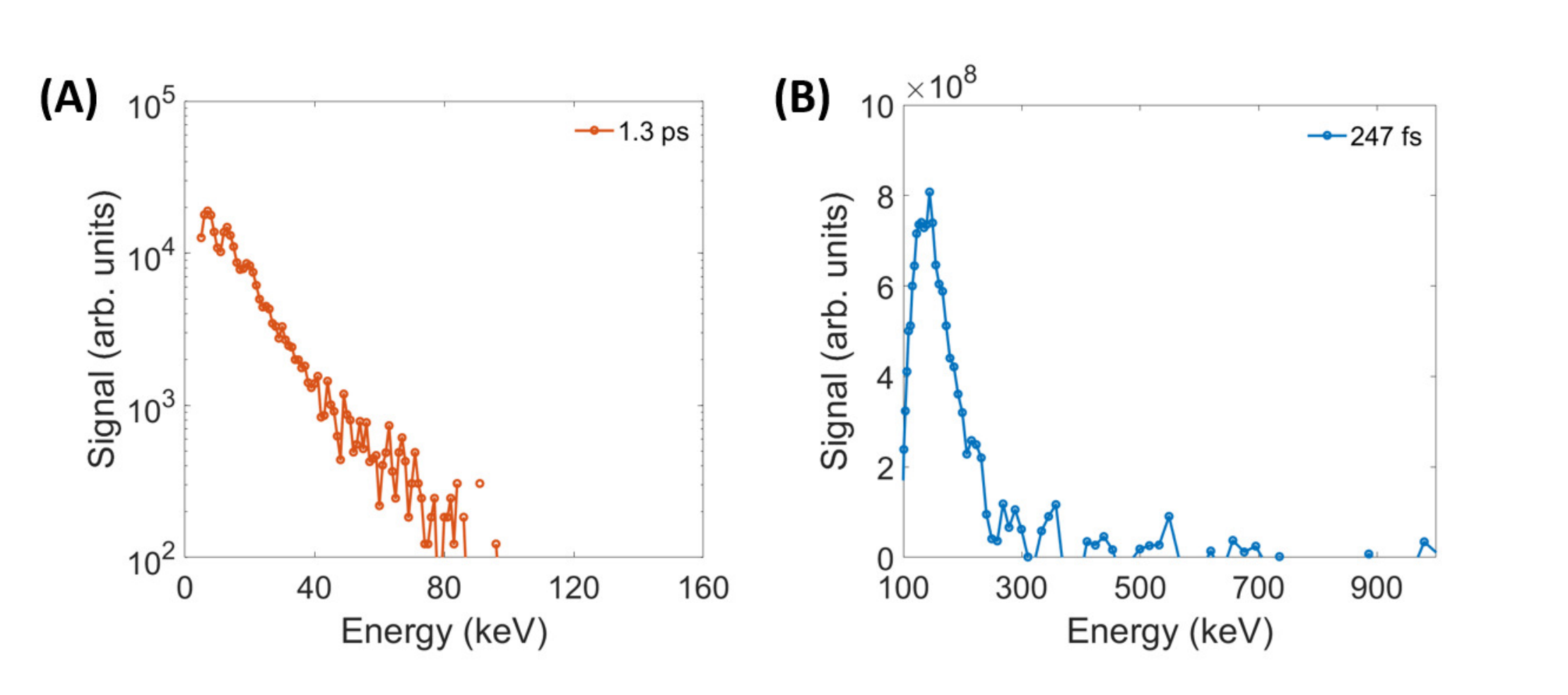}
\caption{(A) X-ray spectrum from the droplet source observed at a stretched pulsewidth of 1.3 ps. (B) Electron spectrum from the droplet source measyred with a 2 mJ, 247 fs pulse.}
\end{figure}

\twocolumngrid